\documentclass[useAMS,usenatbib]{mn2e}
\usepackage{graphicx}
\usepackage{amsmath}
\usepackage{amssymb}
\usepackage{caption}
\usepackage{color}
\usepackage{aas_macros}
\usepackage{url}

\title[Powerful outflows in the LLAGN NGC\,1386]{Powerful outflows in the central parsecs of the low-luminosity Active Galactic Nucleus NGC\,1386}
\author[Rodr\'iguez-Ardila et al.]{A. Rodr\'iguez-Ardila$^{1}$\thanks{E-mail:
aardila@lna.br}; M. A. Prieto$^{2,3}$; X. Mazzalay$^{4}$; J. A. Fern\'andez-Ontiveros$^{2,3}$;
\newauthor R. Luque$^{2,3,5}$; F. M\"uller-S\'anchez$^{6}$\\
$^{1}$Laborat\'orio Nacional de Astrof\'{\i}sica/MCTIC, Rua dos Estados Unidos, 154, Bairro das Na\c c\~oes, Itajub\'a, MG, Brazil\\
$^{2}$Instituto de Astrof\'{\i}sica de Canarias, Calle V\'{\i}a L\'actea, s/n, E-38206, La Laguna, Tenerife, Spain\\
$^{3}$Departamento de Astrof\'{\i}sica, Universidad de La Laguna, E-38205, La Laguna, Tenerife, Spain\\
$^{4}$Max-Planck-Institut f\"ur extraterrestrische Physik, Postfach 1312, D-85741 Garching, Germany\\
$^{5}$Landessternwarte, Zentrum f\"ur Astronomie der Universit\"at Heidelberg, K\"onigstuhl 12, 69117 Heidelberg, Germany\\
$^{6}$Department of Astrophysical and Planetary Sciences, University of Colorado, Boulder, CO 80309, USA
}
\begin{document}

\date{Accepted 2017 June 5. Received 2017 June 5; in original form 2017 February 28}

\pagerange{\pageref{firstpage}--\pageref{lastpage}} \pubyear{2002}

\maketitle

\label{firstpage}

\begin{abstract}
 
Low-luminosity Active Galactic Nuclei, i.e. 
L$_{\rmn bol}$/L$_{\rmn edd} \sim 10^{-6} - 10^{-3} $
constitute the bulk population of Active Galactic Nuclei (AGNs). Powerful jets, common 
in these objects, are a crucial source of feedback energy driving mass outflows into the host galaxy and the
intergalactic medium. This paper reports the first direct measurement of powerful
mass outflows traced by the forbidden high ionization gas in the low luminosity AGN NGC\,1386 at scales of a few
parsecs from the central engine. The high angular resolution of the  data allows
us to directly measure  the  location, morphology and kinematic of the outflow. This
has the form of two symmetrical expanding hot gas shells moving in opposite directions
along the line of sight. The co-spatiality of the gas shells with radio emission
seen at the same parsec scales and with X-rays indicates that this is a
shock-driven outflow presumably induced by an incipient core-jet. 
With a minimum number of assumptions, we derive a mass outflow rate of 11 M$_\odot$ yr$^{-1}$,
comparable to those of  powerful AGN. The result has strong implications in the
global accounting of feedback mass and energy driven by a low-luminosity AGN
into the medium and the corresponding galaxy evolution.

\end{abstract}

\begin{keywords}
Galaxies: seyfert -- Infrared: galaxies -- Galaxies: individual: NGC\,1386
\end{keywords}

\section{Introduction}

Galactic mass outflows are relatively well collimated flows of gas material emerging
out of the plane of a galaxy either from the central regions of starburst galaxies
or from the centre of an active galactic nucleus (AGN). 
They are seen in all gas phases $-$ molecular, neutral and ionized $-$ and are 
identified from the relatively large velocities they reach, varying from a few hundreds up to 
thousands of km\,s$^{-1}$. The energy deposited by
these outflows in their host galaxy, or in  the intergalactic medium if
powerful enough, is a major feedback source  in galaxy formation theories. 
It is currently accepted that these outflows play a major role in the evolution of galaxies, 
stir metals in the Interstellar Medium (ISM), control the common grow of bulges and 
supermassive black holes, 
and possibly halt the star formation in early type galaxies. 

AGN induced outflows are driven by the nuclear radiation,
thermal-pressure,  and/or by jets and are so far rather modest in terms of the 
amount of mass dumped into the medium $-$ few solar masses per year as compared with
hundred to thousand solar masses per year derived form starburst super winds 
\citep[e.g.,][]{heck02+,veill05+}. Unambiguous evidence for AGN
outflows comes  from the detection of blueshifted gas traced in absorption and 
close to the nucleus. That is the case with the reported X-ray, UV and H\,{\sc i}
absorbers \citep[e.g.,][]{crekra05,veill05+,morganti05+}. Still, key parameters to infer the mass outflow rates, such as
density, volume, location and filling factor rely on models, either of the gas
ionization state and/or the outflow morphology. 

Photoionization models are used in the case of the UV/X-rays absorbers to  infer
their location. The results have placed these absorbers at a few tens of parsecs from the AGN, 
possibly in the innermost clouds of the narrow line region \citep[NLR,][]{crekra05}. 
H\,{\sc i} absorbers in radio-loud galaxies are traced at kiloparsecs of distance, possibly 
also associated with the NLR and thus thought to be nuclear outflows \citep{morganti05+}. 
The solid angle subtended at those distances implies a very large mass volume, and thus the mass 
outflow rates derived by these authors  are accordingly among the largest driven by 
AGN, of several tens of solar masses. 

The use of the high ionization coronal line emitting spectrum as a signature of 
energetic AGN-driven outflows at scales of few pc from the centre allowed 
\citet{rod06+} to open a new window to explore this subject. 
Because of their high-ionization potential ($\geq$100~eV), these lines are genuine 
tracers of the presence of an AGN (Penston 1984; Marconi et al. 1994). Their results
show that extended coronal line emission should inevitably involve the presence
of outflowing gas, ionized by the combined effect of radiation from the central
source and shocks formed in the interface of the expanding gas and the ISM. 
Later, \citet{mue11+} employed subarcsec resolution data from adaptive-optics (AO) 
assisted observations from SINFONI \citep{eisen03+}, to model the 2D kinematic 
field of the prominent [Si\,{\sc vi}]~1.96$\mu$m  
coronal line in some of the nearest, best known Seyfert galaxies. They found outflow mass rates
substantially higher than those derived in previous estimates, of a few tens to hundred solar
masses per year. These values depend chiefly on the volume of gas involved, which
is inferred from the model.

Among the closest AGNs displaying a remarkable coronal line spectrum is 
NGC\,1386 \citep{reu02,rod06+}. This source is, instead of a powerful AGN, in the 
borderline between the high luminous Seyfert type and the low 
luminosity AGN (LLAGN) class, as inferred from its low nuclear bolometric luminosity, 
$\sim 3\times10^{42}$ erg\,s$^{-1}$ \citep{fer12}. Its LLAGN nature is further 
supported by the low-Eddington ratio, $\log(L_{\rmn bol}/L_{\rmn edd}) \sim -3.78$ \citep{fer12}.
According to \citet{ho08}, low luminosity AGN display  $\log L_{\rmn bol} \sim 10^{-6} - 10^{-3} $
times the Eddington rate. Very few studies, indeed,
have focused on the high-ionization lines in these type of objects. It is not clear if
the mechanisms leading to the formation of these lines also apply to LLAGN.
It is then of paramount importance to explore the role of such nuclear outflows in
low-luminosity AGN in order to get a clearer picture on the mechanisms powering
the high-ionization lines in them.

Located at 15.3\,Mpc, $1\arcsec \sim 75$\,pc \citep{jen03}, NGC\,1386 is at comparable 
distance as the prototype NGC\,1068. 
Radio observations at different angular resolutions and frequencies confirm that most of the 
radio emission in this galaxy comes from an unresolved source at the centre 
\citep[e.g.,][]{sadler95+,condon98+}. 
The highest angular resolution map available from \citet{mun09} at $8.4\, \rm{GHz}$ shows 
that the radio emission consists of a compact core with an extended jet-like emission at PA = 170$\degr$ 
South of the nucleus. A strong H$_2$O megamaser has also been detected in the nuclear region 
\citep{bwh96+}. 

\begin{figure*}
\includegraphics[width=\textwidth]{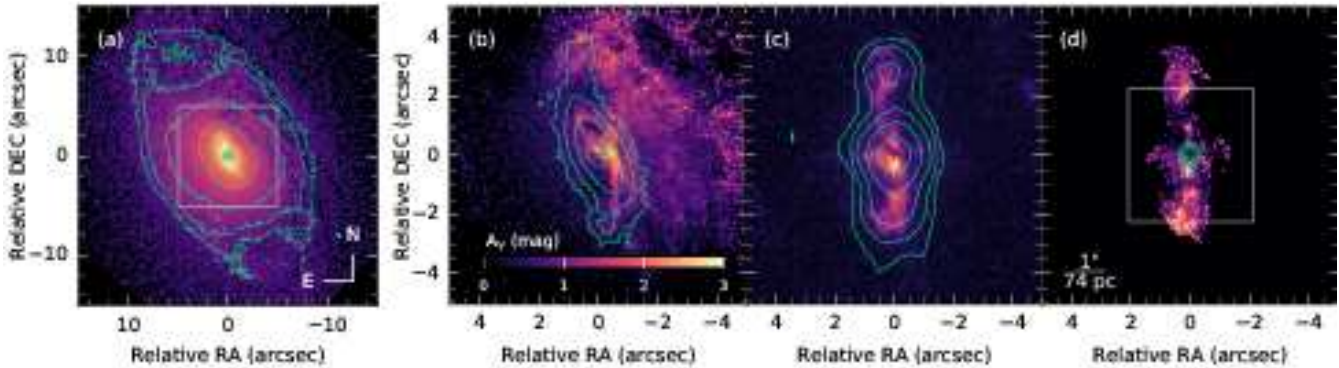}
\caption{(a) {\it HST} broad-band F814W image together with
the NaCo \textit{K}-band image shown in contours. The rectangle shows the zoomed area in 
the following panels. (b) Extinction map in $A_{\rmn V}$ (derived from the \textit{K}/F814W colour map) 
presented in \citet{mez15} together with the H$_2$ narrow-band image from NaCo at 2.121\,$\rmn{\micron}$ in contours. 
(c) H$\alpha$+[N\,{\sc ii}]\,$6548,6584\, \rm{\AA}$ continuum-subtracted image from {\it HST} and soft X-rays 
obtained with {\it Chandra} in contours. (d) Excitation map [O\,{\sc iii}]/(H$\alpha$+[N\,{\sc ii}]) obtained
from narrow-band {\it HST} images together with the radio emission at 8.4\,GHz in contours. The rectangle corresponds
to the SINFONI field of view. In all panels, the ``+'' sign marks the position of the \textit{K}-band nucleus
derived in \citet{pri14}. North is up and East is to the left.
\label{fig:galaxy_overview}}
\end{figure*}

The nucleus of NGC\,1386 shows a wide range of high ionization optical and IR narrow emission-lines. 
In the optical, H$\alpha$+[N\,{\sc ii}]\,$\lambda 6548, 6584$\,\AA\ and 
[O\,{\sc iii}]\,$\lambda$5007~\AA\ 
extend rather collimated along the North$-$South direction up to 6$\arcsec$ from the nucleus, as seen in 
\citet{fwm00} \textit{Hubble Space Telescope} (\textit{HST}) narrow-band images.
\citet{rdw00} resolve the optical line emission in 9 individual components (5 redshifted and 4 blueshifted 
from the systemic velocity). The optical spectra presented in \citet{phfr80,sche03} and \citet{lena15+}  
show the predominance of [N\,{\sc ii}] over H$\alpha$, [N\,{\sc ii}]/H$\alpha$ = 1.5, 
which is in line with a LINER-type (Low-ionization Nuclear Emission Line Region).
The nucleus, though, is classified as a Seyfert\,2 type. 
Spatially resolved emission of [Fe\,{\sc vii}]\,$\lambda$6087~\AA, [Fe\,{\sc x}]\,$\lambda$6374~\AA, and 
[Fe\,{\sc xi}]\,$\lambda$7889~\AA, North and South of the nucleus, are reported by \citet{rod06+}. 
In the 2$\mu$m range, NGC\,1386 shows very strong [Si\,{\sc vi}]\,$1.963\mu$m, 
[Ca\,{\sc viii}]\,$2.321\mu$m and [Si\,{\sc vii}]\,$2.48\mu$m lines (Reunanen et al. 2002). These lines are 
spatially resolved within the central 1$\arcsec$ region. 

The H$_2$ 2.121\,$\mu$m molecular line emission also extends in the North-South direction, 
up to about 4.5$\arcsec$ North, 3.5$\arcsec$ South of the nucleus (Reunanen et al. 2002; 
\citealt{mez15}; this work). Diffraction-limited images at 10\,$\mu$m and 
20\,$\mu$m reveal that the thermal continuum is extended in two well distinct emission blobs 
located at about 0.2$\arcsec$ North and South from the nucleus (Reunanen et al. 2010).

The most accurate estimate of the power of NGC\,1386 comes from the bolometric luminosity
integrated from the high angular resolution SED by 
\citet{fer12} down to scales of few tens of parsec from UV to radio. The SED 
is dominated by the IR bump, and its integration yields a L$_{\rmn bol} = 2.9\times10^{42}$\,erg\,s$^{-1}$, 
about a factor four larger than the previous values. Due to the high angular resolution used, this value 
probably provides the best accounting of the purely reprocessed accretion-disc emission by the central dust 
and put the nucleus in the low-luminosity AGN rank (LLAGN).

The present work analyzes, for the first time in the literature, the central 2$\arcsec$
line emitting properties of NGC\,1386 tracing the molecular H$_2$, the ionized 
Br$\gamma$ and the coronal line gas on spatial scales of 0.1$\arcsec$ ($\sim 8$ pc).  
We give special attention to the kinematic modeling of the molecular and high-ionization
gas as well as to the morphology and extension of the coronal gas and the physical 
mechanisms leading to this emission. To this purpose, this paper is structured as follows. 
In Section~\ref{sec:observations} we describe the observations and data reduction as well
as the most conspicuous features found in the Integral Field Unit (IFU) data.
Section~\ref{stellar} shows the stellar component and its kinematics. The analysis of 
the molecular gas is detailed in Sect.~\ref{sec:mole_kine}. The high-ionization gas, 
its kinematics, emission line properties and the energetics of the outflowing 
coronal gas component are presented in Section~\ref{sec:coronal}. A discussion of the 
overall physical picture emerging from the observations is in Section~\ref{sec:discussion}.
A brief overview of the main results are in Section~\ref{sec:final}. 

\section[]{Observations and data reduction} \label{sec:observations}

Very Large Telescope (VLT)/SINFONI integral-field unit observations of NGC\,1386 
were taken as part of the more extensive PARSEC 
program\footnote{\url{http://www.iac.es/proyecto/parsec/main/group.php}}, which 
is a multiwavelength study of the central 
parsecs of the brightest AGN in the near universe. The VLT/SINFONI data are integral part of the ESO 
observing program 86.B-0484 (PI F. M\''uller-S\'anchez) that was executed in May, 2010. The observations were obtained 
with AO at a spectral resolution of R$ = 4000$, equivalent to $\sigma_{\rmn instr} \approx 35$~km~s$^{-1}$ in the 
\textit{K}-band, and with a pixel scale of $0.10\arcsec \times 0.05\arcsec$. Science
frames were interspersed with sky frames using the sequence Object-Sky-Object to facilitate
background subtraction. The data reduction was performed using the SINFONI pipeline provided 
by ESO and includes: correction for bad pixels, flat-field, geometric distortions, wavelength 
calibration, reconstruction of the data cube from the image spectral slices, background 
subtraction using sky frames and flux calibration using a telluric standard star, also used 
for correction for telluric atmospheric features. Finally, the combined cube
was resampled to a pixel scale of $0.05\arcsec \times 0.05\arcsec$.


\begin{figure*}
\includegraphics[width=130mm]{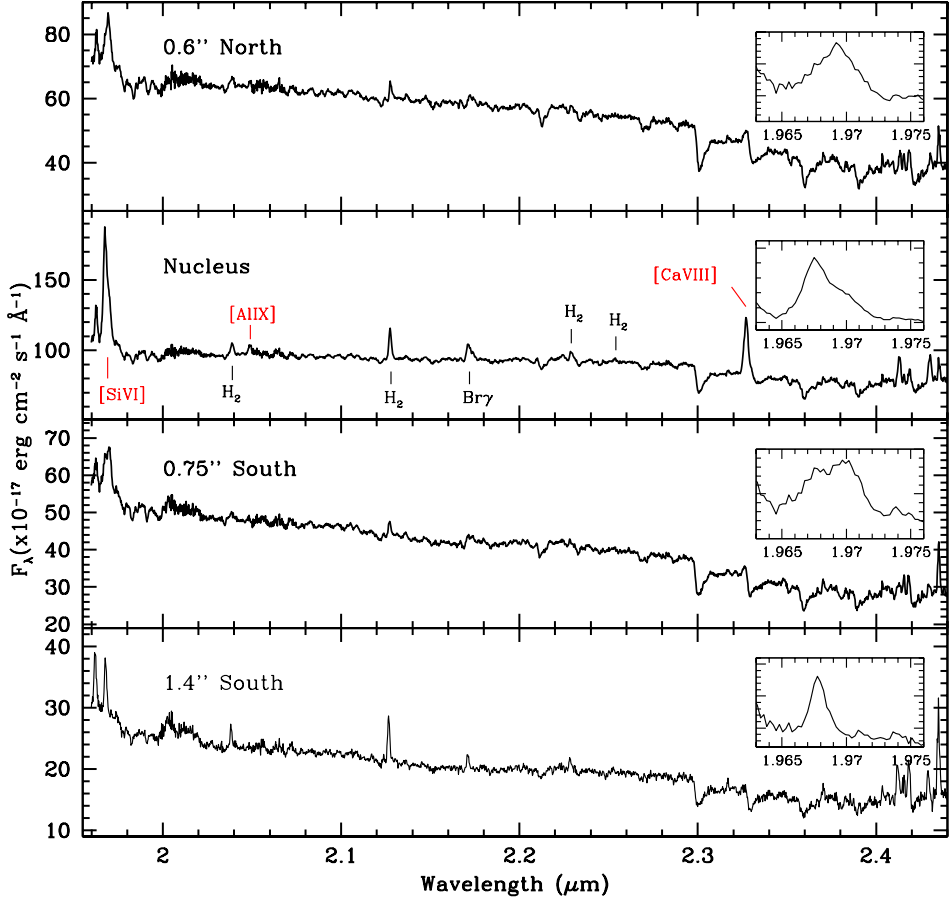}
\caption{Examples of four \textit{K}-band spectra extracted from the IFU cube along the N$-$S
direction in the galaxy frame. In all cases, the integration region covers a circular aperture of 
0.25$\arcsec$ in
diameter. The upper panel shows a spectrum centred at 0.6$\arcsec$ North of the nucleus. 
The second panel (top to bottom) is
the nuclear spectrum. The third and fourth panels show spectra with the aperture centred at 0.75$\arcsec$ and
1.4$\arcsec$, respectively, South of the nucleus. The most conspicuous lines are identified. Coronal
lines are marked in red. The inset in each panel shows a zoom of the [Si\,{\sc vi}]\,1.963$\mu$m
line at that aperture. \label{fig:spectra_overview}}
\end{figure*}

The SINFONI IFU data is complemented with a multiwavelength, high-spatial resolution 
dataset covering the radio, optical, infrared (IR) and X-ray ranges, collected and 
aligned by \citet{fer12}. The dataset includes the 
near-IR VLT/NaCo adaptive optics data in the \textit{K}-band \citep{pri14}. In the 
optical range, high-spatial 
resolution imaging from the {\it HST} scientific archive was taken, including the 
broad-band F547M and F814W filters,
plus the narrow-band F502N and F658N images centred in the [O\,{\sc iii}]\,5007\,\AA\ emission line and the 
H$\alpha$+[N\,{\sc ii}]\,6548, 6584\,\AA\ blend, respectively. Line images were obtained from the narrow-band filters
by subtracting the continuum level based on adjacent broad-band filters, i.e. the [O\,{\sc iii}] continuum was obtained from
F547M and the H$\alpha$+[N\,{\sc ii}] continuum from a linear interpolation between F547M and F814W \citep[as described in][]{pri14}.

The \textit{Chandra} X-ray data were extracted from the scientific  
archive, and analyzed using the \textit{Chandra} Interactive Analysis  
of Observations (\textsc{ciao}) software. The soft X-ray ($0.5$--$2\,  
\rmn{keV}$) image was created using the \textsc{csmooth} task within  
\textsc{ciao} using a minimum significance S/N level of 3. This  
corresponds to the same soft X-ray image presented in \citet{mez15}. 
Only a point-like source associated to the nucleus was  
detected in the hard X-ray band ($2$--$10\, \rmn{keV}$).

In radio wavelengths, Very Large Array (VLA) measurements at $8.4\, \rmn{GHz}$ were also included. These observations 
correspond to the same dataset presented in \citet{mun09}, but were processed and reanalyzed by M. Orienti as part of the 
PARSEC project \citep{ori10}. The radio data were reduced following the standard procedures for the VLA implemented in the 
NRAO \textsc{AIPS} package and tapered to filter the contribution of the longest baselines. The final 
$8.4\, \rmn{GHz}$ image was produced with a circular beam of $0.3\arcsec \times 0.3\arcsec$, following the procedure in 
\citet{mun09}.  The radio and the X-ray nuclear unresolved sources are assumed to be coincident with the position of the 
\textit{K}-band nucleus, since this is the only counterpart found in the FoV.

\citet{pri14} showed that the nucleus of this galaxy has been mistakenly identified with a bright 
\textit{HST}-[O\,{\sc iii}]\,5007\,\AA\ emission region pertaining to the NLR. As already indicated, the true nucleus is 
hidden behind a dust filament of $A_V \approx 2\, \rmn{mag}$.
It is only visible longward of 2\,$\mu$m as an outstanding point-like source. The 
position of the nucleus is $\sim$\,17\,pc North from the optical peak location thought to be the nucleus, or 
0.23$\arcsec$ North as measured from the comparison of IR Adaptive Optics and optical \textit{HST} images used 
in \citet{pri14}. 
The nucleus new location is close to the kinematical centre determined by Schultz and Henkel (2003), 0.6$\arcsec$ 
North East from the optical peak. The larger nuclear shift found by these authors is presumably due to the poor 
angular resolution of their data, $\sim 2 \arcsec$. The identification of the true nucleus affects previous modeling 
and interpretation of the NLR kinematics and nuclear excitation. In this work, we use as a reference the true nucleus location 
identified with the \textit{K}-band continuum peak.

Fig.\,\ref{fig:galaxy_overview} shows a panchromatic view of NGC\,1386: In \ref{fig:galaxy_overview}(a) a wide field of 
view (FOV) image in the F814W filter from {\it HST} together with contours for the NaCo \textit{K}-band continuum (the dust distribution
suggests that the nearest side of the galaxy is in the North-West, while the far side is in the South-East, in an elliptical geometry);
\ref{fig:galaxy_overview}(b) shows the extinction map derived from the \textit{K}/F814W colour map together with the
H$_2$ $2.12\, \rmn{\micron}$ narrow-band image from NaCo by \citet{mez15}; \ref{fig:galaxy_overview}(c) shows the H$\alpha$+[N\,{\sc ii}]
narrow-band image from {\it HST} and the soft X-rays image from {\it Chandra} in contours, showing an extended emission up to
$\sim 4 \arcsec$ in both North and South directions; \ref{fig:galaxy_overview}(d) shows the [O\,{\sc iii}]/(H$\alpha$+[N\,{\sc ii}])
excitation ratio plus the radio emission at $8.4\, \rmn{GHz}$ in contours. In further Sections, the $8.4\, \rmn{GHz}$ radio 
map is shown after subtraction of a $10.8\, \rmn{mJy}$ point-like Gaussian component (FWHM = $0.3\arcsec$) centred in the nucleus, 
in order to reveal the morphology of the extended radio emission.


The high angular resolution \textit{K}-band spectrum of NGC\,1386 shows remarkable coronal lines
(see Figure~\ref{fig:spectra_overview}), with
strong [Si\,{\sc vi}]\,1.963$\mu$m and [Ca\,{\sc viii}]\,2.321\,$\mu$m. [Al\,{\sc ix}]\,2.040\,$\mu$m is also
detected at the nucleus although it is rather faint. The former two dominate over other more common lines in 
the \textit{K}-band spectrum of AGN such as Br$\gamma$ and the H$_2$ molecular emission. 
The spectra also show strong CO band head absorptions. Outside the nuclear region, the silicon and calcium 
coronal lines remain strong and extended over more than 150\,pc distance from the centre.
Figure~\ref{fig:spectra_overview} shows the \textit{K}-band spectrum at various locations, the nucleus and the extended 
coronal line region. The angular resolution achieved in our \textit{K}-band SINFONI spectrum is $< 0.2 \arcsec$
at Full Width at Half Maximum (FWHM), that translates into a spatial scale of $\sim$\,14\,pc.
We have no suitable point-like source in SINFONI field-of-view (FOV), hence we estimate an upper limit to the
resolution taking as a reference the size of the unresolved emission components 
seen in the [Si\,{\sc vi}] gas (see below).

\section{The stellar kinematics} \label{stellar}

An inspection of the spectra plotted in Fig.~\ref{fig:spectra_overview} reveals prominent CO absorption
bands at 2.3\,$\mu$m not only at the nucleus but also in the extended emission. These features
imply the presence of a strong underlying stellar continuum that needs to be removed in order to
obtain a clean emission line spectrum. This is particularly relevant for [Ca\,{\sc viii}]\,2.321\,$\mu$m,
which sits on the edge of one of these bands. In order to 
properly characterize it as well as other emission lines, a clean subtraction of the stellar component
should be carried out. The stellar absorption features also provide a tool to characterize the velocity 
field of the cold galaxy component.

In order to construct the radial velocity map
for the stellar component of NGC\,1386 we used the penalized Pixel-Fitting (pPXF) 
method of \citet{Cappellari+} to fit the stellar continuum. The code
provides the line-of-sight velocity distribution (LOSVD) of
the stars by fitting the stellar absorption features in the interval 2.1--2.37\,$\mu$m.
The regions where the emission lines due to H$_{2}$, Br$\gamma$ and [Ca\,{\sc viii}]
are located were masked.
The pPXF outputs the radial velocity (V$_*$), stellar velocity
dispersion ($\sigma_*$), and higher order Gauss-Hermite moments ($h_{3*}$
and $h_{4*}$), as well as the uncertainties for each parameter.
As stellar templates we used the spectra of
6 late-type stars observed using the same instrumental configuration as the galaxy.

As our main purpose here is to obtain a good representation of the stellar spectrum rather than 
extract information about the stellar population itself, the continuum fit done by pPXF 
allow us to subtract this component. The result is a pure emission line spectra
that will be used in the following sections.

\begin{figure}
\includegraphics[width=\columnwidth]{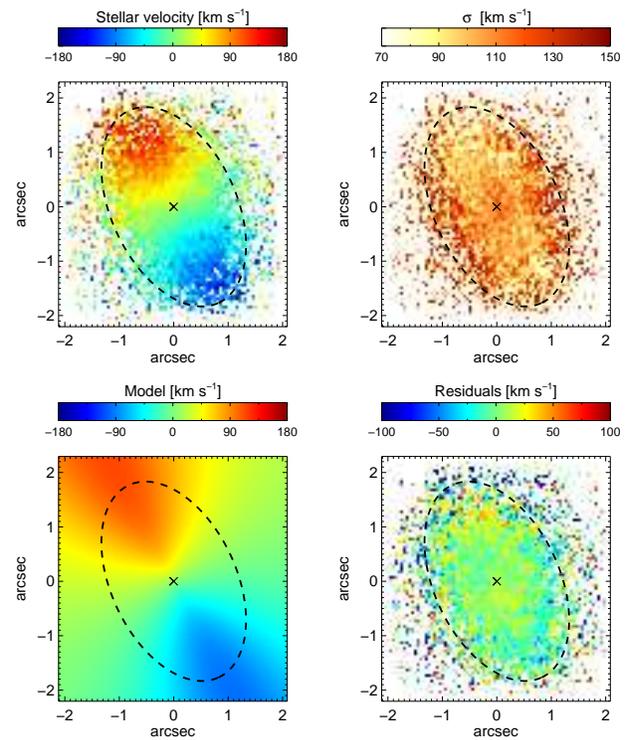}
\caption{\textbf{Top:} Stellar rotation (left) and velocity dispersion (right) as derived from pPXF. 
In all panels, North is up and East is to left. The ellipse in dashed blue marks the points used in the 
fit of the stellar kinematics. Bottom: stellar velocity model (left) and residuals (right) after subtracting 
the model from the observed data.  \label{fig:stellar_kin}}
\end{figure}

Figure~\ref{fig:stellar_kin} shows the velocity map derived for the stellar component
in NGC\,1386 (top left panel) as well as the stellar velocity dispersion ($\sigma_*$,
top right panel). It can be seen that the stellar component displays a clear
rotation pattern in the SINFONI field of view, with the North-East side receding 
from us and the South-West side approaching to us. 

We fit an exponential thin disk model to the points that lie 
inside the ellipse drawn in dashed line in the top left panel of Figure~\ref{fig:stellar_kin}. 
The resulting rotation model and the residuals after subtracting the model from the observations 
are shown in the bottom left and right panels of Figure~\ref{fig:stellar_kin}, respectively.
The small residuals ($\lesssim 30$\,km\,s$^{-1}$) indicate that the stellar component is
dominated by rotation, with a maximum amplitude of $v_{max} \sim$ 130\,km\,s$^{-1}$.
 
The values of the disk inclination 
and position angle (PA) of the major axis found from the fit are $64.7\pm10 \degr$ and $27.8\pm4.2$, 
respectively, in very good agreement to the $I-$band photometry of \citet{Xanthopoulos}.
This suggests that the 
observed circumnuclear disk is very likely aligned with the major axis of the kpc-scale 
galaxy disk.
 
Our values of $\sigma_*$ in the nuclear region ($\sim$130\,km\,s$^{-1}$, see
Figure~\ref{fig:stellar_kin}) are in very 
good agreement to those found by \citet{nelson95+} and \citet{garcia05+} obtained 
by means of long-slit spectroscopy. The former  reported $\sigma$ = $120\pm30$\,km\,s$^{-1}$ 
while the latter found $\sigma_1$ = $123\pm3$\,km\,s$^{-1}$ 
and $\sigma_2$ = $133\pm3$\,km\,s$^{-1}$ using two different methods.

\section{The molecular gas in NGC\,1386} \label{sec:mole_kine}

\subsection{Morphology and kinematics}
The fit done by the pPXF and described in Sect.~\ref{stellar} allowed us to remove 
the underlying stellar continuum to better map the ionic and molecular emission
line gas. We will examine here the transition H$_2$\,2.121\,$\mu$m 1-0S(1), the
brightest molecular line detected in the spectral range covered by the SINFONI
data.

\begin{figure*}
\includegraphics[width=\textwidth]{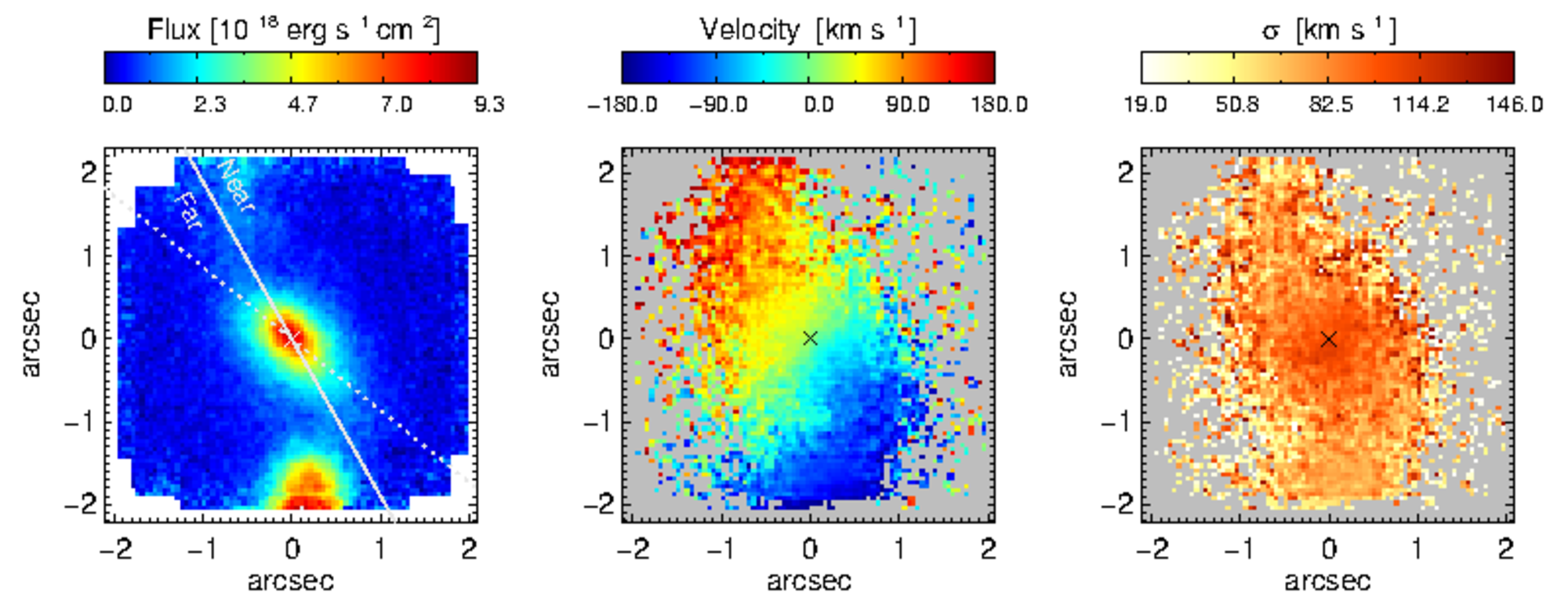}
\caption{Integrated emission line flux in H$_2$\,2.121\,$\mu$m (left). The full 
gray line marks the position angle of the major photometric axis of the galaxy
while the dotted line indicates the PA of the central molecular disk. The labels ``far'' and ``Near''
denote the far and near side of the galaxy.
The central panel shows the observed molecular gas rotation while the right panel displays the
molecular velocity dispersion. In all panels, North is up and East is to left.
\label{fig:h2maps}}
\end{figure*}

In order to extract the emission line flux and kinematics of the molecular gas
we used a Markov chain Monte Carlo (MCMC) technique to fit a Gaussian 
function to the H$_2$\,2.121\,$\mu$m line. From the central wavelength
and width of the Gaussian function we derived the velocity and
velocity dispersion ($\sigma$) of the gas in each spatial bin (0.05$\arcsec$ pixel).

The dust distribution in the central 10$\arcsec$ x 10$\arcsec$ is shown in the 
extinction map of Fig.~\ref{fig:galaxy_overview}b, presented by \citet{mez15}
overlaid to the flux distribution of the molecular hydrogen line. We also indicate 
in the plot the position angle of the major photometric axis of the galaxy 
(full gray line). The dotted line represents the position angle (PA) 
of the central molecular disk ($\theta_{\rmn H2}$ = $50\pm8\degr$, PA = 40$\pm6\degr$ see below). 
It can be seen that the H$_{2}$ emission is asymmetric, with two bright peaks observed. 
One, coincident with the nucleus of the AGN and another at $\sim 2 \arcsec$ south of 
the nucleus. Both regions seem to be connected by a tenuous lane of molecular gas, very prominent
in the dust map distribution. It is also evident in the map an excess of molecular 
emission towards the NE, connected to the nucleus by another lane of dust and molecular gas.

The central panel of Figure~\ref{fig:h2maps} displays the velocity map for 
the H$_2$\,2.121\,$\mu$m. It can be seen that the 
molecular gas displays a global regular rotation pattern across the field, with the NE portion 
of the galaxy rotating away from us and the SW portion approaching us. The largest
velocity measured is $\sim$\,150\,km\,s$^{-1}$.

The right panel of Figure~\ref{fig:h2maps} shows the velocity dispersion $\sigma$ of 
the 2.121\,$\mu$m line across the SINFONI field, already corrected for instrumental 
broadening, subtracting $\sigma_{\rmn inst}$ = 35\,km\,s$^{-1}$ in quadrature.
Regions with very low emission-line fluxes were masked out in the
kinematic maps due to the high uncertainty in the line properties.
These rejected regions, characterized by an emission-line amplitude smaller than five 
times the $rms$ scatter of the continuum, are shown in grey in the 2D maps.
The $\sigma$ measured ranges from 40\,km\,s$^{-1}$ to 140\,km\,s$^{-1}$, with the highest values 
in and around the nucleus, within a region of approximate circular morphology with 
radius $\sim 0.5\arcsec$. The velocity dispersion of the blob at $\sim 2\arcsec$ south
of the nucleus is low, with typical values close to 65\,km\,s$^{-1}$. To
the north, the molecular gas is dominated by $\sigma$ of $\sim$90\,km\,s$^{-1}$. 

Uncertainties for the kinematic parameters were derived by subtracting continua 
equal to the original continuum value $\pm$1 times the continuum noise level. 
This method is more realistic than simply calculating the variance within the 
MCMC samples. The estimated accuracy of the measurements are 
5\,km\,s$^{-1}$ in the case of the velocity of the gas and of 15\,km\,s$^{-1}$ 
for the velocity dispersion.

\begin{figure}
\includegraphics[width=85mm]{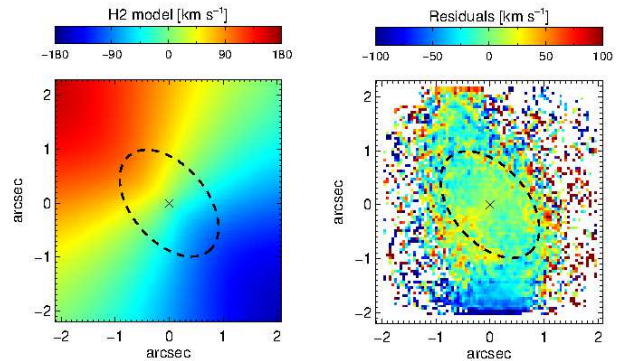}
\caption{Left panel: Best fit model derived from the points located inside the ellipse drawn in gray
from the observed velocity field in Figure~\ref{fig:h2maps} (central panel). 
The fit suggests a disk inclination angle of 50$\pm8\,\degr$ and a PA of the major axis
of 40$\pm6\,\degr$. These values are in good agreement with the ones found from the photometry. Right
panel: residuals left after subtraction of the best-fitting model to the observed velocity field.
\label{fig:h2kin}}
\end{figure}

In order to characterize the rotating velocity field of the molecular gas, we fitted a
rotating exponential thin-disc model to the observed velocity map following 
the procedure described in \citet{maz14}. The points considered in the fit were those
located along and within the ellipse drawn in dashed line in  Fig.~\ref{fig:h2kin}.
The resulting velocity field is shown in the middle panel of Fig.~\ref{fig:h2kin}. The model
that best describes the observations is a disk at an inclination of $\theta$ = 50$\pm8\degr$ 
with the line of nodes at PA = 40$\pm6\degr$ (dotted line in the left panel of Fig.~\ref{fig:h2kin}). 

The results above show that the model that best describes the kinematics of the molecular gas 
component slightly diverges from the one derived from the stellar 
population (see Sect.~\ref{stellar}). Although, within errors, the disk inclination agrees to 
that of the stellar population, the position angle of both disks
vary by more than 10$\degr$. This indicates that both components may be misaligned 
in the central few tens of parsec of this object. The kinematical decoupling of both disks 
is interpreted in the literature as evidence of a bar, a nuclear spiral or a stochastic inflow due 
to external accretion into the central kiloparsec of a galaxy \citep{davis14+}. 
In order to distinguish between these scenarios, observations with better S/N in the 
H$_2$ line is necessary.

The subtraction of the rotation model to the data (see the right panel in Fig.~\ref{fig:h2kin}) 
shows that the bulk of the molecular gas is rotation-dominated. Residuals as high as 40\,km\,s$^{-1}$ 
both in blueshift and redshift are observed at some positions. Our results here overall 
agree with those presented by \citet{hicks13+},
who shows that the kinematics of the molecular gas in Seyfert galaxies is dominated by rotation rather
than bulk outflow and is thus a reasonable tracer of the cool ISM.  

\subsection{Excitation of the molecular gas} \label{sec:h2gas}

The detection of four rotation-vibrational transitions of the H$_2$ molecule in the SINFONI spectrum 
(Fig.\,\ref{fig:spectra_overview}) allows us to study the origin of the molecular gas excitation.
H$_2$ emitting gas can be excited by three mechanisms: (i) fluorescent excitation through
absorption of soft-UV photons in the Lyman and Werner bands \citep{bla87};
(ii) collisional excitation due to the heating of the gas by the
interaction of a radio jet with the interstellar medium \citep{hol89};
or (iii) X-ray illumination, where X-ray photons penetrate deep into molecular clouds,
warming up the interior to temperatures suitable to produce warm H$_2$ emission. 
Which of these mechanisms dominate can be studied through the flux ratio of the 
H$_2$ 2-1 S(1) 2.247\,$\mu$m and 1-0 S(1) 2.121\,$\mu$m lines in the 
{\it K}-band \citep{reu02,rod04,dor12}.

The line fluxes for 2-1 S(1) and 1-0 S(1) transitions were extracted
using apertures centred on the nucleus and on the southern extension.
In order to estimate the continuum and background contributions, a measurement 
with the same aperture radius was extracted from an offset position within the 
SINFONI FOV, in a region without line emission. The latter was subtracted
from the nuclear and southern extension measurements, to provide the final flux 
values.

In our case, the ratio of the 2-1 S(1) 2.247\,$\mu$m and 
1-0 S(1) 2.121\,$\mu$m lines is 0.14 for the
nuclear region and 0.12 for the southern extension. The values obtained for the 
1-0 S(2) 2.033\,$\mu$m and
1-0 S(0) 2.223\,$\mu$m ratio were 1.95 and 1.97, respectively.
According to the diagnostic diagrams presented by \citet{dor12}, these line ratios
suggest X-ray illumination as the main excitation mechanism for the molecular
gas emission. This result agrees with that
of \citet{mal96}, \citet{dor12} 
and \citet{mez15}, who found that heating by X-ray emission seems
to be the most important mechanism of H$_2$ gas excitation in active galaxies.
Note that the models do not distinguish the source of X-rays. 
In the case of an AGN, it is natural to consider the central engine as 
the main souce of high-ionization radiation. However, shocks between the radio-jet
and the ISM can also produce extended X-ray emission and drive the observed
molecular emission.

Figures\,\ref{fig:galaxy_overview}(b)~and~(c) show the distribution of the molecular gas 
traced by H$_2$ 2.121\,$\mu$m emission and the soft 0.5--2\,keV X-rays image taken 
from {\it Chandra}, respectively. 
In this case, the NaCo dataset was used because of its larger FOV, thus the whole extent 
of the molecular  gas and the X-rays can be covered. The X-rays, the ionized gas traced by 
H$\alpha$+[N\,{\sc ii}], and the molecular gas are completely co-spatial, although the 
molecular gas shows a warped morphology, possibly due to the rotation. 
The similar morphology of these extended components further supports the H$_2$ being due to X-ray heating. 
In Sect.~\ref{phot_shock}, we argue that shock excitation is necessary to explain the ionization of 
[Si\,{\sc vi}] seen at $> 100$\,pc from this AGN. Shock velocities greater
than 200\,km\,s$^{-1}$ produce strong free-free X-ray emission and thus can be the 
cause of the observed extended X-rays and, in turn, additionally contribute to the H$_2$ excitation.

\section{The coronal-line gas} \label{sec:coronal}

Previous works in the optical range have already shown that NGC\,1386
displays strong coronal lines of [Fe\,{\sc vii}]\,$\lambda$5721\,\AA, 
[Fe\,{\sc vii}]\,$\lambda$6087\,\AA\ and [Fe\,{\sc x}]\,$\lambda$6374\,\AA\
\citep{rdw00,bennert06+,rod06+}.
Moreover, evidence of an extended coronal emission was reported by \citet{rdw00} and
\citet{rod06+}, who found that [Fe\,{\sc vii}]\,$\lambda$6087\,\AA\ is emitted
in a region of $\sim$\,100\,pc in radius from the the centre. \citet{rod06+} 
also report that [Fe\,{\sc vii}] 
displays a double-peaked profile, with the red peak stronger than the blue one. This
result was interpreted in terms of an outflow of high-ionization gas, with the
blue and red peaks representing the approaching and receding components,
of the outflow, respectively. This scenario was confirmed recently by 
\citet{lena15+} by means of GEMINI/GMOS IFU spectra (see below).

\subsection{Morphology and kinematics}

\begin{figure*}
\includegraphics[width=130mm]{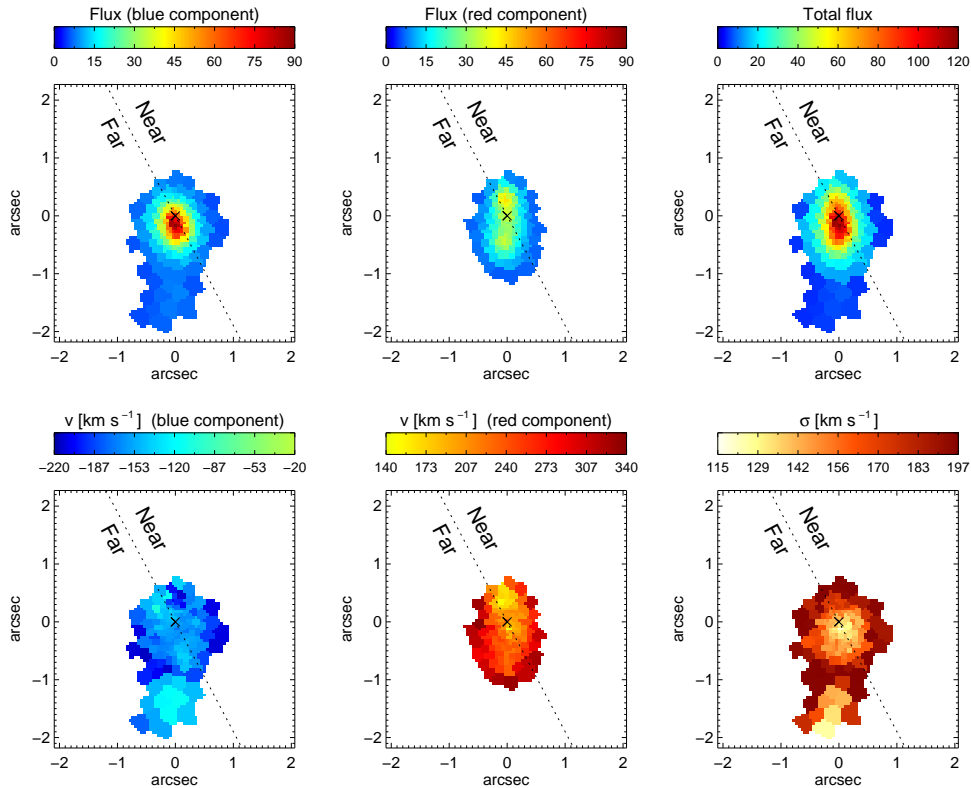}
\caption{Integrated emission line flux distribution for [Si\,{\sc vi}]\,1.963\,$\mu$m in units
of 10$^{-17}$\,erg\,cm$^{-2}$\,s$^{-1}$. The
top left and middle panels present the results for the blue and red components, respectively.
The right panel shows the total flux measured in that line. The bottom panels show the 
velocity map of the blue and red components (left and central panels, respectively), and 
the velocity dispersion of the observed profile (right panel) The dashed line corresponds to 
the photometric PA of the galaxy; the far and near sides of the galaxy are indicated. 
In all panels, North is up and East is to left.
\label{fig:si6maps}}
\end{figure*}

The excellent signal-to-noise (S/N) of the VLT/SINFONI datacube and wavelength 
coverage of the \textit{K}-band spectra allow us to map, for the first time 
in the literature for this AGN, the morphology 
and kinematics of the coronal lines [Si\,{\sc vi}]~1.963~$\mu$m,
[Ca\,{\sc viii}]~2.32~$\mu$m and [Al\,{\sc ix}]~2.04~$\mu$m. Figure.~\ref{fig:si6maps} shows the 
flux map constructed for the former line in the central 4$\arcsec \times 4\arcsec$ of NGC\,1386. 
It can be seen that [Si\,{\sc vi}] is not symmetrically distributed across the FOV, displaying 
two prominent regions of emission. One, characterized by a blob of $\sim 1\arcsec$ in radius, 
centred at the AGN and slightly elongated in the N$-$S direction. The brightest region of this 
component is highly elongated in the N$-$S direction, 
with a size of $\sim 1.2\arcsec \times 0.6\arcsec$.
A second region of emission is visible to the south, starting from the southern rim of
the blob and extending up to 2$\arcsec$, very close to the southern edge of the FOV. 

[Ca\,{\sc viii}] is dominated by the blob already 
seen in [Si\,{\sc vi}], with similar size and morphology found in the later. 
For this reason, it will not be shown here.
The strong tail of emission detected in silicon, though, is not observed in calcium. 
[Al\,{\sc ix}] is highly concentrated, covering a region of size $\sim 0.3\arcsec$. 

An inspection to the coronal line profiles observed in the central blob of emission reveals
complex features, with most emission lines splitted in at least two components. This
is illustrated in Fig.~\ref{fig:si6maps}, which show the decomposition into blue and
red components carried out to the [Si\,{\sc vi}] line. The upper panels show the flux
distribution of the blue component (left), red component (centre) and total flux (right).
The gas that contributes most to the blue component is located in the central circular
blob and along the extended emission to the south (seen only in [Si\,{\sc vi}]). The region emitting
the redshifted component is clearly elongated in the N$-$S direction and spatially restricted to the blob. 
No hint of this emission is observed at distances larger than 1.1$\arcsec$ to the south in none 
of these two lines. In [Si\,{\sc vi}], the inner portion of the region
emitting the red component has two bright spots. The strongest one is centred at 0.3$\arcsec$ north 
of the nucleus while the secondary spot is at 0.45$\arcsec$ to the south, both connected by a faint 
bridge of gas emission.

The central blob of emission seen in [Si\,{\sc vi}] and [Ca\,{\sc viii}] is very similar in 
size and morphology to the one seen in the [Fe\,{\sc vii}]\,6087\,\AA\ \citep{lena15+}. 
They also report splitted lines, with the blue and red components distributed in a region of circular 
shape. Because of the higher angular resolution in SINFONI observations, we are able to 
disentangle physical structures not shown in their maps. Note that the extended emission to 
the south, seen in the blue peak of [Si\,{\sc vi}] is not visible at all in the 
[Fe\,{\sc vii}]\,6087\,\AA\ map.

The velocity distribution of the blue and red components of the [Si\,{\sc vi}] line, shown in 
the left and central bottom panels of Figure~\ref{fig:si6maps}, 
indicate that the relative shift between the red and blue peaks remains nearly constant
along the region where they are detected. On average, a separation of 
$\sim$\,350\,km\,s$^{-1}$ is measured, with random variations of $\sim$\,30\,km\,s$^{-1}$
at some positions. That separation in velocity is very similar to the one observed
in the blue and red peaks in the [Fe\,{\sc vii}] line \citep{lena15+}. It suggest that the gas
emitting the NIR coronal lines is co-spatial to that producing the optical iron lines.

The gas velocity dispersion $\sigma$ for [Si\,{\sc vi}] (bottom right panel of 
Figure~\ref{fig:si6maps}) shows a minimum at the nuclear region, with $\sigma \sim 140$\,km\,s$^{-1}$.
The velocity increases radially with distance, reaching values close to $\sigma \sim 200$\,km\,s$^{-1}$
at the outermost border of the central blob. This same effect is also seen in [Ca\,{\sc viii}]
in spite of the lower S/N of this line. In [Si\,{\sc vi}], the tail to the south is characterized by low
velocities, of $\sigma < 140$\,km\,s$^{-1}$.

\begin{figure}
\centering
\includegraphics[width=80mm]{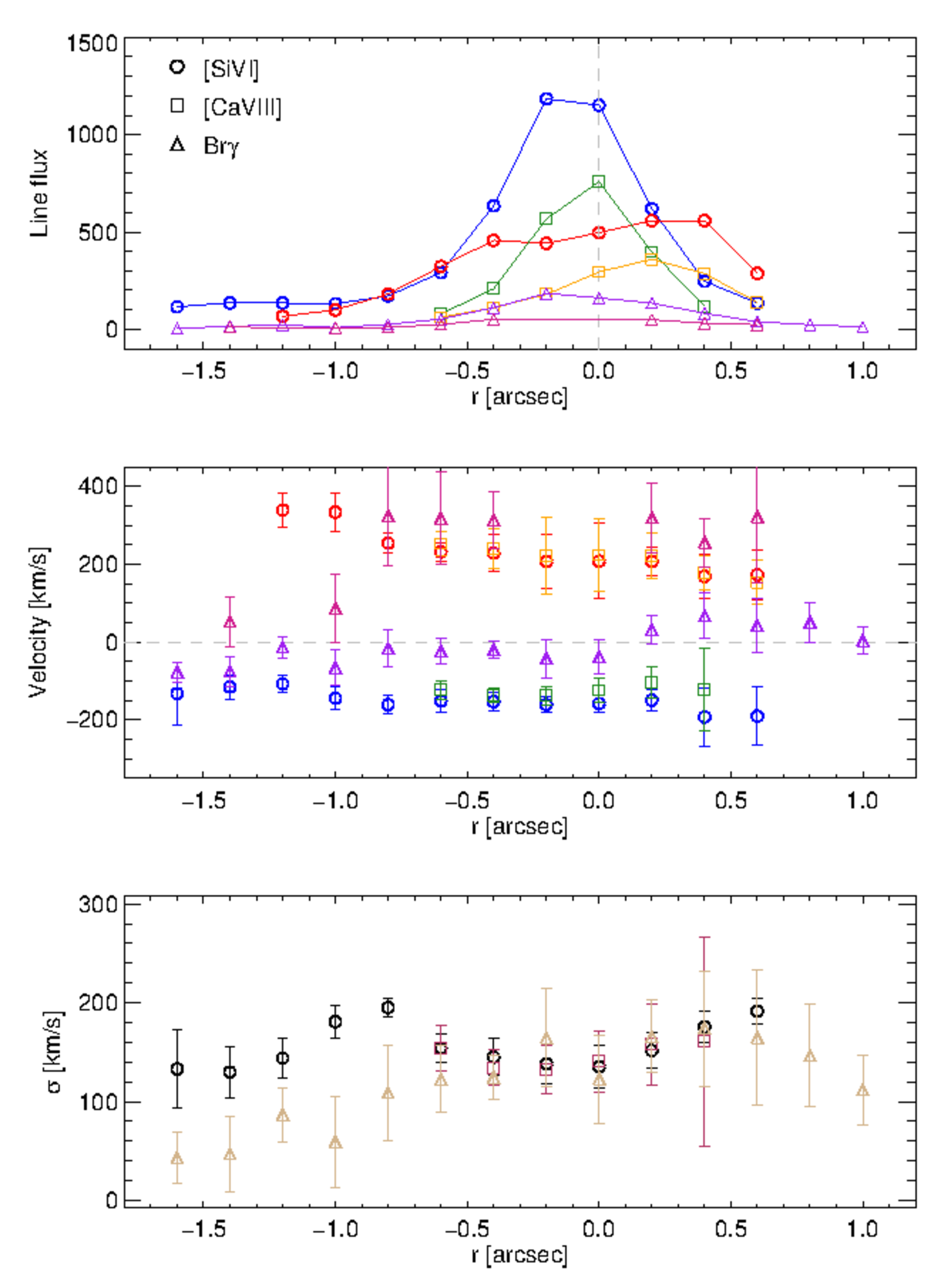}
\caption{Flux distribution (top panel, in units of 10$^{-17}$\,erg\,cm$^{-2}$\,s$^{-1}$), peak velocity (middle panel) and velocity dispersion (bottom panel)
for [Si\,{\sc vi}] (circles), [Ca\,{\sc viii}] (squares) and Br$\gamma$ (triangles) measured 
in consecutive regions of $0.2\arcsec \times 0.2\arcsec$ along the N-S direction. Negative/positive radii 
correspond to distances south/north from the nucleus. For lines with splitted components, 
we report the results for the blue and red peak at the corresponding positions. 
\label{fig:2g_longslit}}
\end{figure}

In order to better visualize the relative strength, relative peak separation and velocity dispersion
of the different ionic lines detected in NGC\,1386, we have extracted 1-D spectra at steps of 0.2$\arcsec$ 
along a pseudo-slit of 0.2$\arcsec$ wide, crossing the AGN and oriented in the N$-$S direction. 
The results are plotted in Fig.~\ref{fig:2g_longslit}. North 
is to the right and south to the left of the position marked as ``zero'', which coincides with the 
position of the true nucleus. In the upper panel we see the integrated flux of [Si\,{\sc vi}] (circles), 
[Ca\,{\sc viii}] (squares) and Br$\gamma$ (triangles). For the silicon line, the blue and red curves 
represent the blue and red components, respectively. It can be seen that the flux distribution 
of each component is different, with the former being rather cuspy, displaying a nearly symmetrical
distribution in the inner 1$\arcsec$. The flux distribution of the red component has a shallow shape 
along the region where it is detected. In the inner 1$\arcsec$, a small intensity gradient in
flux, increasing from south to north, is apparent. 

The light profile of the blue and red components of [Ca\,{\sc viii}] (green and orange curves, 
respectively) follow very closely their counterparts in the [Si\,{\sc vi}] line except for the 
lack of extended emission to the south, as already mentioned. Br$\gamma$, in contrast, 
behaves rather different, with a shallow light profile both in the redshifted and
blueshifted components. 

The second panel of Fig.~\ref{fig:2g_longslit} displays the velocity curve obtained 
from the blue and red components of [Si\,{\sc vi}], [Ca\,{\sc viii}] and Br$\gamma$. 
Two results are evident after inspection of this plot. First,
the coronal gas in NGC\,1386 is not rotation-dominated, in contrast to what
is observed for the stellar and molecular gas, in agreement with what is often found for
the kinematics of coronal line gas \citep{mazzalay10+,maz13b,mue11+}. 
Second, the high-ionization gas 
lies predominantly out of the plane defined by the the molecular gas. Moreover,
the constant velocity shift with distance between the approaching and receding components of the 
silicon and calcium lines suggests shells of gas expanding in opposite directions, with the 
main axis of expansion nearly aligned with the line-of-sight to the observer.  

Our results strongly contrast to that found by Lena et~al. (2015) for the emission gas, represented
by the [N\,{\sc ii}]\,$\lambda$6584\,\AA\ line, the strongest emission line they analyzed. 
The velocity map presented for that line shows the characteristic pattern of a
rotating disk upon which other components are superposed. They also found, from the velocity dispersion 
map, that the strongest line broadening occurs in the nuclear region and is roughly centered on the 
nucleus. Figs.~\ref{fig:si6maps} and~\ref{fig:2g_longslit} show exactly the 
opposite: in the nuclear region, the coronal gas displays the narrowest profiles while the
broadest lines are found at $\sim 0.5 \arcsec$ from the AGN. Note, however, that Br$\gamma$
shows a different trend to that of the coronal gas but yet different to that of [N\,{\sc ii}].
Overall, the line profiles broaden when going from south to north, with a local minimum coincident
with the location of the AGN. Further to the north after crossing the nucleus, the gas velocity
dispersion behaves randomly. Note that the error bars are large, though.

\subsection{Channel maps Coronal lines and low-ionization gas} 

\begin{figure*}
\includegraphics[width=130mm]{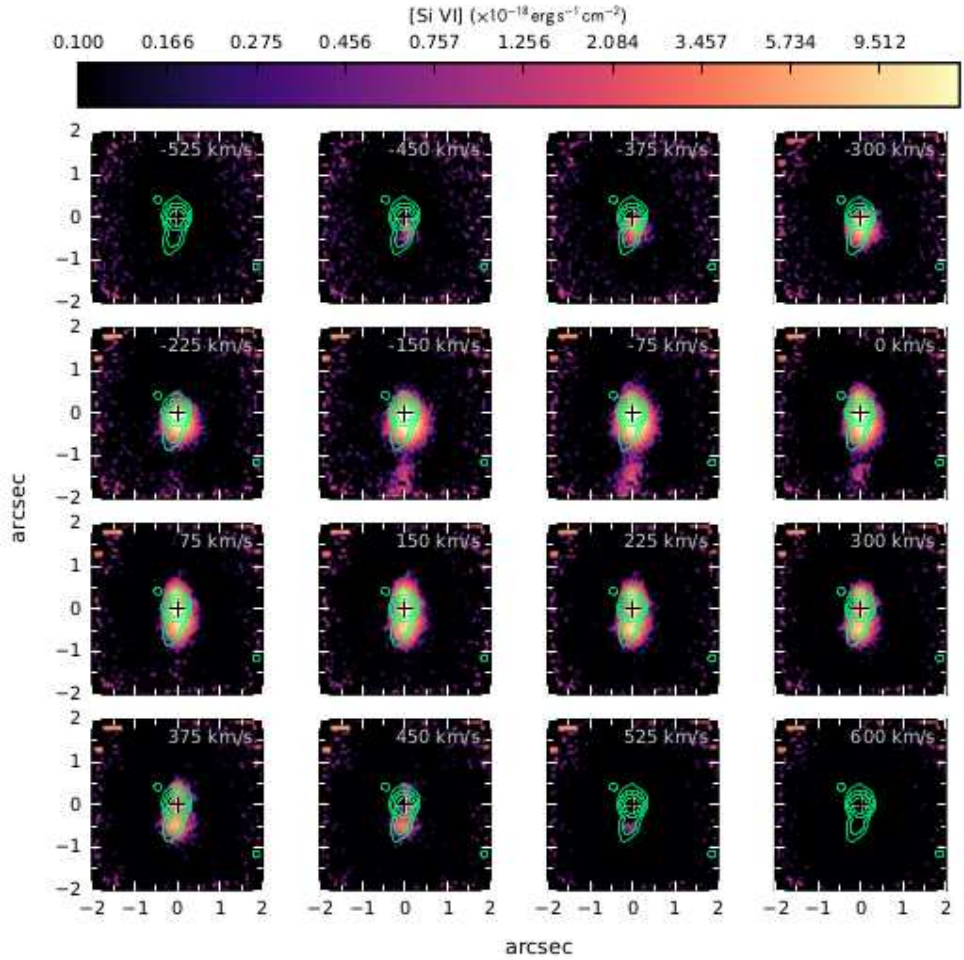}
\caption{
Channel maps derived for the [Si\,{\sc vi}]$1.963\, \rmn{\micron}$ coronal line. A velocity 
bin of $75\, \rmn{km\,s^{-1}}$ is used to slice the datacube. The corresponding velocity is shown
in the upper right corner. The green contours correspond to the $8.4\, \rmn{GHz}$ radio
data, after subtraction of the nuclear unresolved source. In all panels, the ``+'' sign 
marks the position of the \textit{K}-band nucleus derived in \citet{pri14}. North is up and 
East is to the left. \label{fig:s6channel}}  
\end{figure*}

\begin{figure*}
\includegraphics[width=130mm]{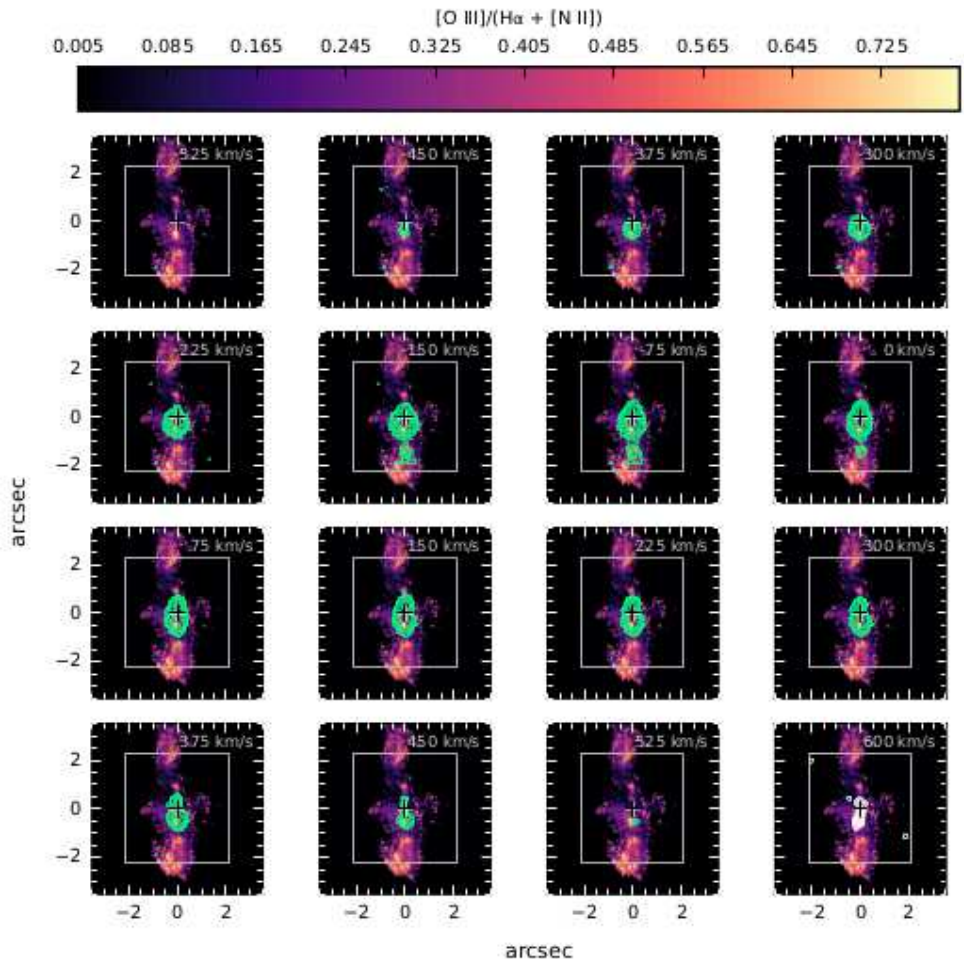}
\caption{Channel maps derived for the [Si\,{\sc vi}]\,1.963$\mu$m line (contours) 
overlaid to the excitation ratio [O\,{\sc iii}]/H$\alpha$+[N\,{\sc ii}] corresponding to the 
integrated [O\,{\sc iii}] emission from WFPC2/\textit{HST} imaging described in 
Ferruit et al. (2000). A velocity bin of 75\,km\,s$^{-1}$ is employed between two consecutive 
panels. The corresponding velocity is shown in the upper right corner. The white box in each panel
represents the FOV of the VLT/SINFONI. In all panels, the ``+'' sign marks the position of the \textit{K}-band
nucleus derived in \citet{pri14}. North is up and East is to the left. \label{fig:s6channelO3}}
\end{figure*}

\begin{figure*}
\includegraphics[width=130mm]{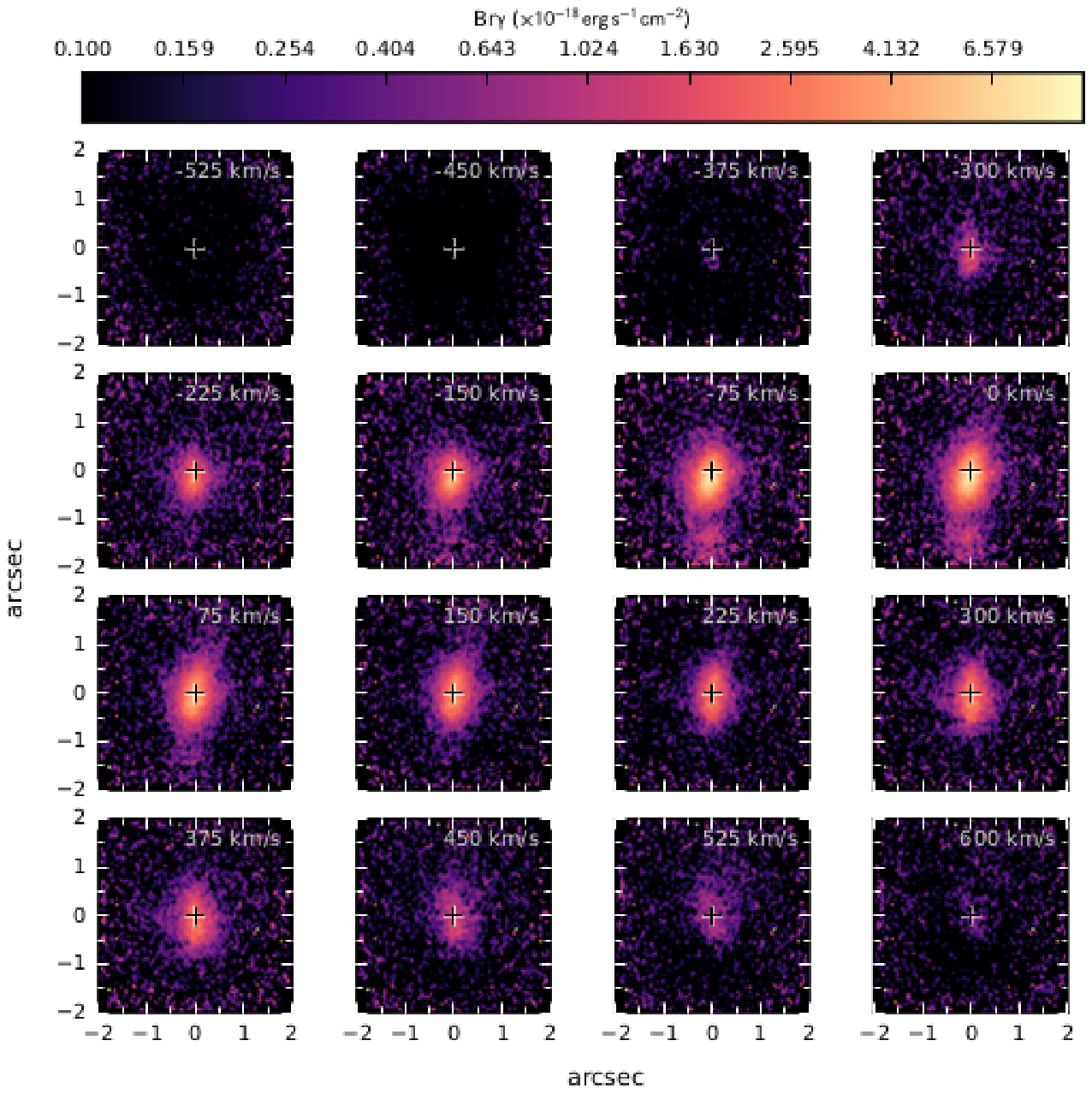}
\caption{
Channel maps derived for the Br$\gamma$~2.166~$\mu$m line. A velocity 
bin of $75\, \rmn{km\,s^{-1}}$ is used to slice the datacube. The corresponding velocity is shown
in the upper right corner. In all panels, the ``+'' sign marks the position of the \textit{K}-band
nucleus derived in \citet{pri14}. North is up and East is to the left.} \label{fig:brgchannel}
\end{figure*}

\begin{figure*}
\includegraphics[width=130mm]{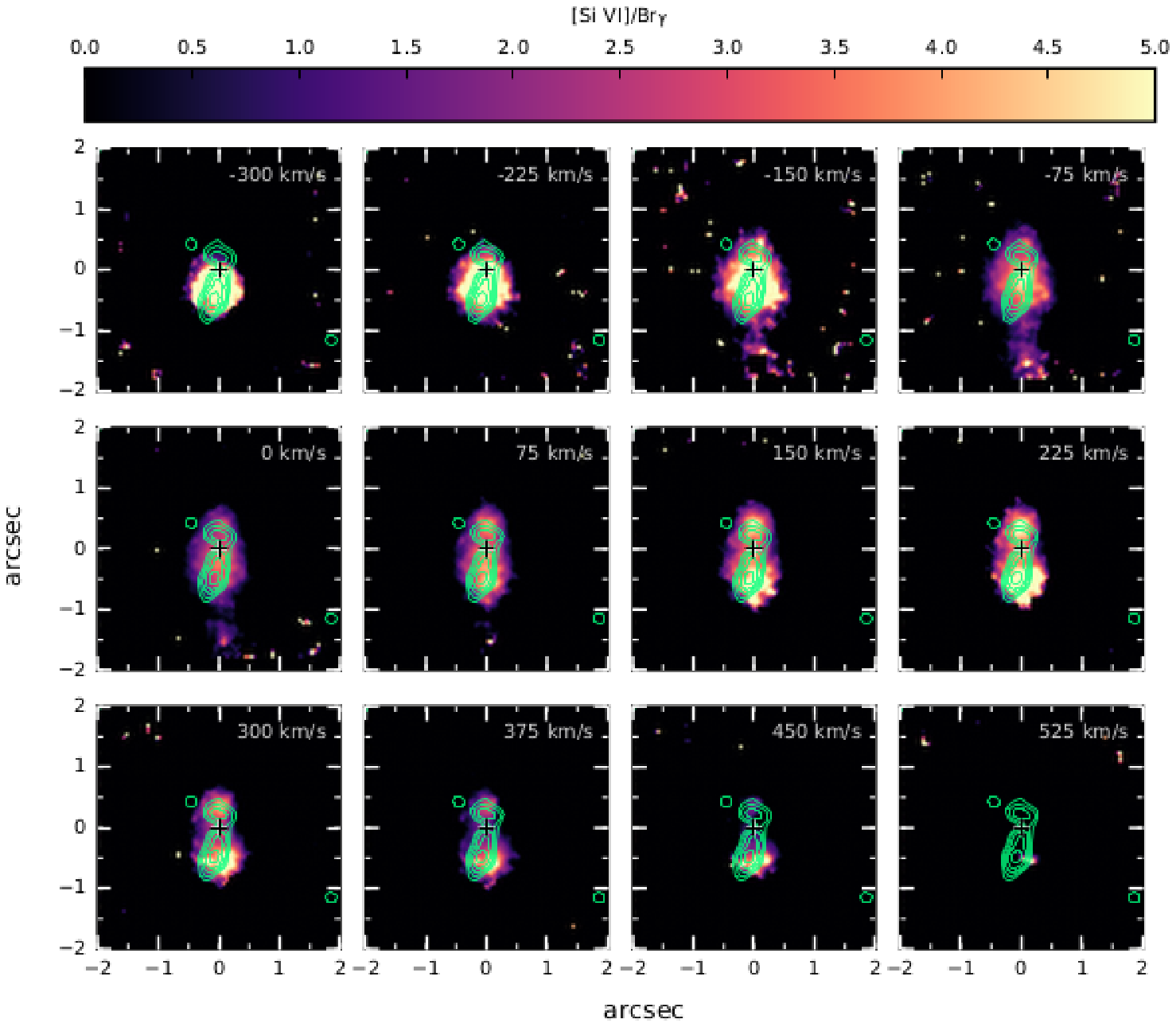}
\caption{Channel maps derived for the excitation ratio [Si\,{\sc vi}]/Br$\gamma$ (value range shown in the upper colourbar). 
A velocity bin of $75\, \rmn{km\,s^{-1}}$ is employed between two consecutive panels. The corresponding velocity is shown in 
the upper right corner. Green contours correspond to the $8.4\, \rm{GHz}$ radio data, after subtraction of the nuclear 
unresolved source. In all panels, the ``+'' sign marks the position of the \textit{K}-band nucleus derived in \citet{pri14}. 
North is up and East is to the left.} \label{fig:s6_brg}
\end{figure*}

Additional information for gas the kinematics can be obtained by slicing the observed 
line profiles in velocity bins. The location where highly blueshifted or redshifted gas
is produced can be mapped through this technique. With this in mind, channel maps for 
the coronal line [Si\,{\sc vi}] as 
well as for Br$\gamma$ were constructed after subtracting the systemic velocity of 
the galaxy. The results are shown in Figs~\ref{fig:s6channel} to~\ref{fig:s6_brg}.

For [Si\,{\sc vi}] (Fig~\ref{fig:s6channel}), a velocity bin of 75\,km\,s$^{-1}$ was
employed. Overlaid to each velocity channel are contours of the $8.4\, \rmn{GHz}$ radio
data, after subtraction of the nuclear unresolved source (see Sect.~2). We were able to map the observed 
[Si\,{\sc vi}] from the very blue end, at -525\,km\,s$^{-1}$, to the red end at 600\,km\,s$^{-1}$. 
An inspection to Fig~\ref{fig:s6channel} shows that the parcel of gas associated to the
redshifted component follows closely the distribution of the radio emission, from the systemic velocity
up to 450\,km~s$^{-1}$. Indeed, at  this latter velocity, the ionized gas is mostly concentrated 
at the position coincident with the secondary radio peak south of the nucleus. The blueshifted coronal
gas, on the other hand, is distributed in a nearly circular region, whose centre is
offset relative to the centre of the radio emission, and in the extended
tongue to the south. We interpret this result in terms of two opposite shells of gas along 
the line of sight, one approaching and the other moving away from the observer.

Channel maps for [Ca\,{\sc viii}] are very similar
to those of [Si\,{\sc vi}] except that the extended emission to the south is absent. For that
reason they are not shown here. The lack of the southern blob in [Ca\,{\sc viii}]
implies that the region emitting this line is rather compact. We also notice the lack
of high blueshifted and redshifted gas emitting that line compared to that of the silicon. 
However, the relative peak separation between the blue and red peaks is similar
to that of silicon in the common region where both lines are observed. This points out that 
the gas producing these NIR coronal lines (and very likely, the optical ones) is
co-spatial. Note that the lack of high-velocity components in [Ca\,{\sc viii}] may reflect
the lower S/N around this line relative to that of [Si\,{\sc vi}] and not due to
intrinsic reasons.

Figure~\ref{fig:s6channelO3} shows the channel maps of Fig~\ref{fig:s6channel} for 
[Si\,{\sc vi}] but now overlaid to the excitation ratio [O\,{\sc iii}]/H$\alpha$+[N\,{\sc ii}] 
from WFPC2/\textit{HST} imaging. A close look to the images reveal that the coronal gas 
follows closely the spatial distribution displayed by the optical emission, although the
former is much more compact. Some regions strongly affected by
dust extinction in the optical are now nicely traced by the silicon gas, particularly those located
to the West and North$-$West side of the galaxy nucleus. Figure ~\ref{fig:s6channelO3} 
also shows the excellent correspondence
between the optical ionization map and the coronal gas. For instance, the hot spot located
0.5$\arcsec$ south of the centre is clearly noticed in the [Si\,{\sc vi}] channel maps between 
225\,km~s$^{-1}$ and 525\,km~s$^{-1}$. Moreover, the southern tongue of [Si\,{\sc vi}] 
seen in the blue component, from $v >$ -224\,km\,s$^{-1}$ up to
the systemic velocity, coincides with the system of blobs of emission seen in 
[O\,{\sc iii}]/H$\alpha$+[N\,{\sc ii}] map. 

Channel maps for the Br$\gamma$ line are shown in Figure~\ref{fig:brgchannel}.
A velocity bin of 75\,km\,s$^{-1}$ was employed. It can be seen that this line
is significantly narrower than [Si\,{\sc vi}] as it extends just from -300\,km~s$^{-1}$
to -400\,km~s$^{-1}$. Moreover, the morphology of the emission region from
-70\,km~s$^{-1}$ to -130\,km~s$^{-1}$ is ellipsoidal, with its major axis oriented
in the NE$-$SW direction. This contrast with what is seen in the coronal lines, with 
the emission region either circular for the blue-peak or elongated to the south in the
red peak. Br$\gamma$ is also intrinsically fainter than [Si\,{\sc vi}],
mainly at the highest velocity bins, both in negative and positive velocities.
This is seen in Figure~\ref{fig:s6_brg}, where the line ratio [Si\,{\sc vi}]/Br$\gamma$
is observed. This map can also be regarded as an ionization map, similar to that
shown in Figure~\ref{fig:s6channelO3}. However, because both lines are equally affected
by extinction, the effect of dust is cancelled out here.

Figure~\ref{fig:s6_brg} confirms the results already seen above. Three main excitation
regions are clearly noticed in NGC\,1386. The first one is associated
to the AGN, where the presence of bright, broad wings in the [Si\,{\sc vi}] line produces
[Si\,{\sc vi}]/Br$\gamma$ ratios greater than 5. The second region is associated to the
secondary peak of the radio-emission, at $\sim 0.5 \arcsec$ South of the AGN, where both 
[Si\,{\sc vi}]/Br$\gamma$ and [O\,{\sc iii}]/H$\alpha$+[N\,{\sc ii}] reach 
line ratios larger than 7 and 0.8, respectively. This hot spot is more prominent in the
velocity bins $v >$ 225\,km~s$^{-1}$. Indeed, at this later velocity, several substructures
are organized around the southern tip of the radio emission, suggesting shock
excitation of the high-ionization lines. The third region corresponds to the extended
tongue of [Si\,{\sc vi}] South of the AGN, observed up to the edge of the IFU FOV.
It coincides in position with a bright spot seen in the 
[O\,{\sc iii}]/H$\alpha$+[N\,{\sc ii}] map.

\subsection{Photoionization by the central source vs. shocks}\label{phot_shock}

Observations of AGNs using IFU/AO \citep{mue11+,maz13} have revealed extended 
coronal line emission up to scales of a few hundred of parsecs. This high-ionization 
gas is usually aligned to the radio-jet, arranged in multiple knots of emission. 
The complexity of the line profiles, with splitted lines, and the kinematics
inferred from the data show that the coronal gas is non-rotating, most probably 
outflowing. This result contrasts to previous long-slit, seeing-limited 
observations of AGNs, where a central unresolved coronal gas emission was usually 
detected, leading to the picture where the these lines were formed between the 
NLR and the BLR \citep{erkens97+,rod02+}. 

\citet{rod06+} were the first to observationally resolve the coronal emission by 
means of optical and NIR observations. NGC\,1386, one of the 
AGNs of their sample, displayed [Fe\,{\sc vii}] emission with splitted profiles, 
not observed in low$-$ and mid$-$ionization lines such as [O\,{\sc i}]
and [O\,{\sc iii}], respectively. This, along to the high 
values of flux line ratios between coronal lines led them to propose that the bulk 
of the coronal gas was in an outflow, with the line emission powered by a combination 
of photoionization by the central source and local shocks produced in the interface 
of the outflowing gas and the ISM. The excellent angular resolution achieved here 
with the IFU/SINFONI data allows us to investigate more 
deeply these previous results since we are now able to resolve coronal gas 
emission at scales down to $\sim 0.2\arcsec$.

In order to investigate whether photoionisation by the central source
can be responsible for the extended coronal line emission,  we
generated a grid of models using \textsc{cloudy} \citep[Version C13.03][]{ferland13+}. 
Our main interest here is to find out if the observed luminosity
of NGC\,1386 is able to power high-ionization gas at distances as large as 140\,pc
from the centre. We are also interested in comparing the observed line flux ratios
to those predicted by the models. Due to the reduced set of ionic lines detected
in the spectral range covered by SINFONI, a robust fit to the whole coronal line region
(CLR) spectrum is out of the scope of this paper. We may, however, derive general physical properties
of the coronal line gas by combining NIR data with that collected in the optical
range.

The input to the models include the gas density, $n_{\rm H}$; the distance 
of the clouds to the nucleus, $R$; the AGN luminosity, the spectral energy
distribution of the ionizing radiation, the elemental abundances,
the dust/gas ratio, and the column density of the
emission-line clouds. Solar abundances from \citet{grevesse10}
were employed in all cases. The numerical abundances relative to hydrogen are as
follows: He = 8.51 $\times 10^{-2}$,
C = 2.69 $\times 10^{-4}$, O = 4.9 $\times10^{-4}$, N = 6.76 $\times 10^{-5}$, 
Ne = 8.51 $\times 10^{-5}$, S = 1.32 $\times 10^{-5}$, Si = 3.24$\times 10^{-5}$, 
Mg = 3.98 $\times 10^{-5}$, and Fe = 3.16$\times 10^{-5}$. No other values of 
gas metallicity were used as no reliable indicator of this variable exists 
using NIR lines. 

The ionizing continuum employed was similar to that deduced by \citet{mafer87}. It is meant 
to represent a typical radio-quiet AGN continuum and consists of several broken power-laws of the
form $f_{\nu} \propto \nu^{\alpha}$ with $\alpha$ taking different values according to the
wavelength range. The intrinsic luminosity of NGC\,1386 above the 
Lyman limit was set to 2.9 $\times 10^{42}$ erg\,s$^{-1}$ (Sect. 1). Clouds with 
densities $n_{\rm H} $=500, $n_{\rm H} $=$ 10^{3}$ and $n_{\rm H} $=$10^{4}$ cm$^{-3}$ 
were considered at distances $R$ from 0 to 120\,pc, which is within the range of the observed 
extent of the [Si\,{\sc vi}] and Br$\gamma$ emission.

\begin{figure}
\includegraphics[width=\columnwidth]{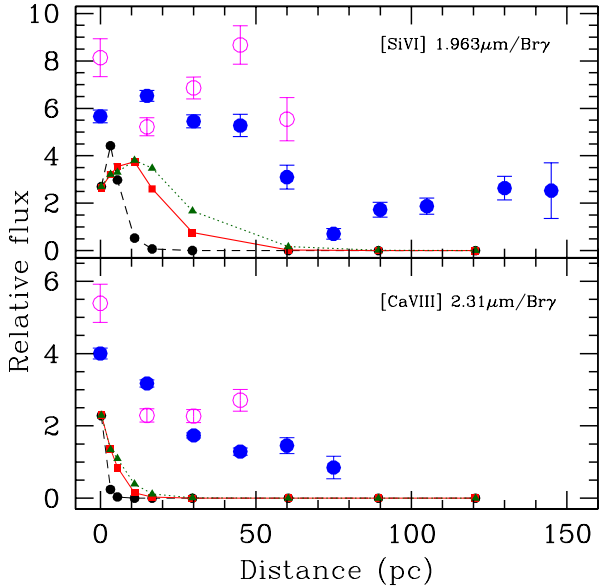}
\caption{Predicted emission line rations [Si\,{\sc vi}]/Br$\gamma$ (upper panel) and
[Ca\,{\sc viii}]/Br$\gamma$ (bottom panel) for clouds of density $n_{\rm H} $=500 (full triangles), 
$n_{\rm H} $=$ 10^{3}$ (full squares) and $n_{\rm H} $=$10^{4}$ cm$^{-3}$ (full black circles).
The full circles are the observed line ratios for the blue component of the profiles while the
open circles represent those of the red profile. \label{fig:cloudy}}
\end{figure}

Fig.~\ref{fig:cloudy} shows \textsc{cloudy} outputs for [Si\,{\sc vi}] (upper panel) and
[Ca\,{\sc viii}] (bottom panel) relative to Br$\gamma$ for the three densities considered. 
Open circles correspond to positions North of the nucleus and full circles
to those located to the South. These positions correspond to centroid position of the
1D spectra extracted
along the pseudo-slit. From the figure,
it is evident that no coronal emission can be produced with the assumed set of input parameters
at distances larger than 50~pc. Indeed, if the density of the coronal gas is 
$n_{\rm H} \geq 10^{4}$ cm$^{-3}$, the silicon emission region is restricted 
to the inner 20\,pc. Only clouds with $n_{\rm H} \leq 10^{3}$ cm$^{-3}$
are able to create a larger emission region, which however, is not larger than 80\,pc
away from the nucleus. Such low coronal gas densities, though, are unlikely to occur in this
AGN. For example, \citet{fer16} using the  \textit{Spitzer}/IRS spectrum for NGC\,1386, 
derived a value of $n_{\rm H} = 10^{3}$ cm$^{-3}$
using the [Ne\,{\sc v}] lines in this galaxy. Note also that the emission distribution 
for all clouds fall out
steeply with distance. This strongly contrasts to the observations.
Model predictions for [Ca\,{\sc viii}] (bottom panel of Figure~\ref{fig:cloudy}) 
also do not reproduce the VLT/SINFONI data.
We see that the emission region for that ion is even more compact than that for [Si\,{\sc vi}]. 
The inclusion of low density clouds ($n_{\rm H} \leq 10^{3}$ cm$^{-3}$) does not improve the fit 
as the size of the emission region is still restricted to the inner 20\,pc. We are able to detect 
[Ca\,{\sc viii}] from regions located at least 3 times farther away.

Another important drawback of the models is that under the physical conditions assumed,
they are unable to produce [Fe\,{\sc x}] and [Fe\,{\sc xi}], even at the nucleus. These lines 
were clearly detected by \citet{rod06+} by means of optical spectroscopy for
NGC\,1386. Moreover, [Al\,{\sc ix}], a NIR line detected from our data is also 
not predicted by the pure photoionization models. It can be argued that the coronal line
gas is illuminated by a harder continuum that the one we see. We tested this possibility by computing 
models where the central source is 10, 100 and 1000 times more luminous than the fiducial value
adopted while keeping the other parameters fixed. The results show that NGC\,1386 would 
need to be at least 100 times more luminous to
solve the caveat of producing higher line flux ratios to match the observations.
However, even under this scenario, the [Si\,{\sc vii}] region does not extend beyond 80\,pc, in 
a clear disagreement with the observations.

Photoinization by the central source alone
may explain the lines ratios (within a factor of 2x) [Si\,{\sc vi}]/Br$\gamma$
and [Ca\,{\sc viii}]/Br$\gamma$ at the nucleus and in 
the inner few tens of parsecs away. However, at distances larger than 50\,pc,
model outputs largely underpredicts our observations. This result suggests that additional processes
should be present at these locations to enhance the high-ionization spectrum. 
The characteristic double-peak line profiles seen in the ionized
gas can be taken as kinematic evidence for 
outflows in the NLR of NGC\,1386. The fact that the extended emission runs aligned to
the observed radio-jet suggests that part of the gas ionization may be directly influenced
by that component. Thus, it makes sense to consider the role of shocks produced by interactions
between the radio jet or a radially accelerated outflow and
the ISM. \citet{rod06+} had already
found strong evidence of shocks powering the coronal gas emission in NGC\,1386 
based on the observed ratios between the optical coronal lines [Fe\,{\sc vii}], 
[Fe\,{\sc x}] and [Fe\,{\sc xi}]. 

The models of \citet[][hereafter CV01]{Contini+}
are useful to test the scenario of shocks coupled
to photoionization by the central source; even in the presence of shocks, the effect of the
central continuum cannot be ignored. In their modelling, CV01
considered that the clouds are moving outwards from the galaxy centre, 
with the shock front forming on the outer edge of the cloud, whereas 
the ionizing radiation reaches the opposite edge that faces the active centre. 
The ionization due to both the primary radiation (from the central source)
and the diffuse radiation generated by the free-free and free-bound
transitions of the shocked and photoionized gas, as well as the collisional
ionization, are all accounted for. The shock velocity $V_{\rm s}$ and
the ionizing flux from the central source at the Lyman limit reaching
the cloud, $F_{\rm h}$ (in units of cm$^{-2}$\,s$^{-1}$\,eV$^{-1}$), are the main 
input parameters. Other parameters include the pre-shock density, $n_0$, and
the pre-shock magnetic field, $B_0$.

Tables~1 to~12 of CV01 show that shock-dominated clouds ($F_{\rm h}$ = 0) with shock 
velocities in the range 300 $-$ 500 km\,s$^{-1}$ strongly favour the production of 
[Si\,{\sc vi}]. They predict [Si\,{\sc vi}]/Br$\gamma$] ratios between 13 and 2.1, 
for $V_{\rm s}$ = 300 and 500\,km\,s$^{-1}$, respectively. In these models, solar 
metallicity, $n_0$ = 300\,cm$^{-3}$ and B$_0$ = 10$^{-4}$ Gauss were adopted. When 
coupled to the presence of the radiation field from the central source (e.g. log $F_{\rm h}$ = 12, 
models 62 and 83), the value of that ratio, for the same two velocities above,
are $\sim$\,0.2 and $\sim$\,1.7, respectively). These values are rather consistent with our 
observations. They match the measured line flux ratio of $\sim$\,2 for [Si\,{\sc vi}]/Br$\gamma$
in the extended region to the South.
Because shocks strongly impact the gas ionization locally, its effects are independent
to the distance from the central source. In regions where photoionization by the central
source alone is not able to produce coronal lines, shocks are a possibility that should be
considered. We found strong support to this scenario, and for the first time 
find consistent proofs that the well-defined extended coronal emission to the south 
is powered by this mechanism. Note that we cannot discuss the results for [Ca\,{\sc viii}] 
because that line is not available in CV01's results. 

We are aware that the shock model coupled to radiation from 
the central source predictions provide us with only a first-order
approximation. A full description of the CL region based
on model-fitting is far beyond the scope of this paper.
However, our data provide us with solid evidence of the necessity of shocks 
coupled to photoinization by the AGN in
order to reproduce CL flux ratios at distances
as far as 120\,pc, where photoionization by
the central source predicts no or very faint CL emission. 
The superior angular and spectral resolution of the VLT/SINFONI 
observations confirms the previous results by \citet{rod06+}
for the presence of shocks in NGC\,1386, based on an independent 
set of coronal lines. Unrealistic 
physical conditions need to be assumed for the gas if we want to
explain the observed high-ionization lines in terms of
only one mechanism.


\subsection{AGN-driven outflows traced by the coronal gas} \label{subsec:feedback}

We use the velocity and the angularly resolved morphology of the outflowing [Si\,{\sc vi}] coronal gas to derive the mass 
outflow rate and the mechanical power inserted in the ISM. Assuming a shell morphology of $\sim 1.5\arcsec \times 0.5\arcsec$ ($\sim 111 \times 37\, \rm{pc^2}$), 
an outflow velocity of $v_{\rm out} \approx 225\, \rm{km\,s^{-1}}$ with a velocity dispersion of $\sigma \approx 150\, \rm{km\,s^{-1}}$, 
the mass outflow rate and the mechanical power can be derived in the following way:
\begin{equation}\label{eq:energy}
\begin{aligned}
  \dot{M}_{\rm out} = & 5.5 \left( \frac{m_{\rm p}}{1.67 \times 10^{-24}\, \rm{g}}\right) \left(\frac{n_{\rm e}}{10^{2.97}\,{\rm cm^{-3}}}\right) \\
  & \times \left(\frac{l}{111\,{\rm pc}}\right) \left(\frac{w}{37\,{\rm pc}}\right) \left(\frac{f}{0.1}\right) \left(\frac{v_{\rm out}}{225\,{\rm km\,s^{-1}}}\right)\, \rm{M_\odot\, yr^{-1}} \\
  \dot{E}_{\rm kin} = & \frac{1}{2}\dot{M}_{\rm out} (v^2_{\rm out} + \sigma^2) \sim \\
  & 8.3\times10^{40} \left[ \left(\frac{v_{\rm out}}{225\, \rm{km\,s^{-1}}}\right)^2 + \left(\frac{\sigma}{150\,{km\,s^{-1}}} \right)^2 \right]\, {\rm erg\,s^{-1}},
\end{aligned} 
\end{equation}
    where $m_{\rm p}$ corresponds to the proton mass, $l$ and $w$ refer to the length and the width of the blob, respectively. 
    A density of $10^{2.97}\, \rm{cm^{-3}}$ (error interval $10^{2.83}$--$10^{3.08}\, \rm{cm^{-3}}$) was derived by \citet{fer16} from 
    the [Ne\,{\sc v}]\,24.3$\mu$m/[Ne\,{\sc v}]\,14.3$\mu$m line ratio measured with {\it Spitzer}/IRS in 
    the high-spectral resolution mode. The largest uncertainties in the mass outflow rate and the mechanical power estimates come 
    from the gas filling factor ($f$), widely discussed in the literature. \citet{mue13} assume $f = 0.001$ based on the relation 
    $n_{\rm e} \propto f^{-1/2}$ found by \citet{oli97} for the coronal line gas. However, theoretical and observational
    arguments presented by  \citet{sha10} showed that for ionized gas winds in nearby AGN and starburst galaxies
    a higher value of $f \sim 0.1$ is needed. They, indeed, found out that $f$ is likely distributed over the interval
    $0.1 - 1$. This result is in agreement with previous estimates made by \citet{blufa09}. They used constraints imposed by 
    the observed radio emission to obtain upper limits to the volume filling factors of wind components in nearby AGN.
    Alternatively, if the outflow is considered as a violent blast of gas, as in the case of supernovae explosions, higher $f$ 
    values may be expected for the hot coronal gas \citep[$f \lesssim 0.6$, e.g.][]{li15}. For this work, we adopt a conservative 
    value of $f \sim 0.1$.

With the above assumptions, the outflow mass rate of coronal gas derived from Eq.\,\ref{eq:energy} is $\sim 5.5\,{\rm M_{\odot}\,yr^{-1}}$, 
a total of $\sim 11\, \rm{M_{\odot}\,yr^{-1}}$ if we consider both the redshifted and the blueshifted components. The associated kinetic 
power of the redshifted component is $\sim 8.3 \times 10^{40}\, \rm{erg\,s^{-1}}$, and a total of $\dot{E}_{\rm kin} \sim 1.7 \times 10^{41}\, \rm{erg\,s^{-1}}$ 
including also the blueshifted component, which is $\approx 6\%$ of the total radiative energy ($2.9 \times 10^{42}\, \rm{erg\,s^{-1}}$).
An upper limit to the kinetic energy can be obtained from the terminal velocity observed in the channel maps of 
[Si\,{\sc vi}] (Fig.\,\ref{fig:s6channel}), $\sim 450\, \rm{km\,s^{-1}}$, which results in $\dot{E}_{\rm kin} < 10^{42}\, \rm{erg\,s^{-1}}$. 
For a radio core luminosity of $L \sim 10^{36.1}\, \rm{erg\,s^{-1}}$ at $5\, \rm{GHz}$ (derived from $10.8\, \rm{mJy}$ at $8.4\, \rm{GHz}$ and 
$\alpha = 0.47$, $S_\nu \propto \nu^\alpha$, from \citealt{fer12}), the kinetic outflow energy measured for the coronal gas would be 
in agreement with an extrapolation of the \citet{mer07} relation at low luminosities.

\subsection{The inflow rate traced by H$_2$}

The total molecular gas mass available was estimated from the H$_2$\,2.121~$\mu$m 1-0 S(1) luminosity. 
Determining the total gas mass requires first the conversion of H$_2$ luminosity to warm H$_2$ gas mass and second, the conversion 
from warm to cold gas mass. Here we use the conversion from \citet{maz13}:
\begin{equation}
 M_{\rm cold_{\rm H_2}} \approx 1174 \times \frac{L_{1-0\,S(1)}}{\rm L_{\odot}}\ {\rm M_{\odot}} \sim 2\times10^7\ {\rm M_{\odot}} ,
\end{equation}
where $L_{1-0\,S(1)}$ is the luminosity of the H$_2$\,2.121~$\mu$m emission integrated in the apertures 
described in Sect.~\ref{sec:h2gas} using a distance of 15.3\,Mpc \citep{jen03}.
Figure~\ref{fig:h2kin} suggests some systematic residuals from rotation, albeit in the noise level, 
of $\sim\,30$\,km\,s$^{-1}$. 
Taking this figure as an upper limit to any radial motion in molecular gas, assuming a density of $n \sim 10^4$\,cm$^{-3}$ \citep{dor12}, and a 
filling factor of $\sim 0.1$; we derive an upper limit for the inflow mass rate of $< 0.4\,{\rm M_{\odot}\,yr^{-1}}$ using Eq.~\ref{eq:energy}.


\section{Discussion}\label{sec:discussion}

In this section we will envisionage the most plausible scenario for the ionized
gas in NGC\,1386 based on the observed morphology, kinematics and the flux 
distribution presented in the previous sections. 
We have already shown that the morphology of the high and low ionization gas in the central 100~pc is 
preferentially elongated towards North and South of the nucleus with a somewhat peanut shape.
The kinematics of the coronal gas shares the same 
velocity amplitude, $\pm$ 200~km\,s$^{-1}$, regardless of its location relative to the nucleus. 
The simplest explanation to this configuration is that of a pair of nuclear outflowing shells, 
one getting away from us $-$ with positive velocities $-$ and  its counterpart moving  towards us  
with negative velocities. Effectively, we see one on top each other as projected on the sky and 
across the nucleus, so that any point in the shell in the  N$-$S direction across the nucleus has about 
the same velocity, $\sim$ 200~km\,s$^{-1}$, for the receding shell and $\sim$175~km\,s$^{-1}$, 
the approaching one (see Fig.~\ref{fig:2g_longslit}).

These expanding shells, which are about bisected by the nucleus, are presumably  driven by the AGN 
radiation pressure or mechanically by a jet. NGC\,1386 is a low Eddington source, 
$\log(L_{\rm bol}/L_{\rm edd}) \sim -3.78$ \citep{fer12}, and thus radiation-driven winds are not expected to be 
its dominant output energy channel \citep[e.g.][]{schartmann14+}. Low Eddington sources 
are often powerful radio sources, best prototype is M87 \citep[][and references therein]{prieto16+}, 
and thus a jet in NGC\,1386 is a plausible cause of the outflowing shells.
NGC\,1386 shows extended radio emission at each side of the nucleus along the N-S direction, 
the Southern emission being better defined and detached-from-the-core while the northern one 
appears as a protuberance. The peak of emission to the South in the positive velocity bins 
shows a remarkable coincidence in morphology and location in the [Si\,{\sc vi}], 
[Si\,{\sc vi}]/Br$\gamma$ and  [O\,{\sc iii}]/H$\alpha$ channel maps (see Figs.~\ref{fig:s6channel} 
to~\ref{fig:s6_brg}). The similarity  in morphology leads us to conclude that 
the extended radio emission is due to synchrotron emission  
generated locally downstream the front shock created between the jet and the ISM 
by e.g. the Fermi mechanism.  An alternative  N-S two-side jet scenario may still produce 
the receding and approaching gas shells by pushing material laterally in directions 
perpendicular to the  jet orientation. However, it may not be expected in this case the jet 
morphology to coincide with that of the gas. A second problem is that the jet component 
pointing towards us, which in this scenario would be the Southern extended radio emission, 
would be expected to coincide with the blueshifted rather than the redshifted gas.

We propose that a core jet $-$ coinciding with the central radio source (Sect.~2) $-$ moving 
in the line of sight causes respective front shocks that we identify with 
the receding and approaching  shells. The jet kinetic power 
transferred to the gas yields gas bulk velocities of up to 450~km\,s$^{-1}$. Main  locations where 
the energy transfer occurs are identified at $\sim$30 $-$ 40~pc from the centre coinciding with  
the maxima in [Si\,{\sc vi}]/Br$\gamma$, [O\,{\sc iii}]/H$\alpha$+[N\,{\sc ii}],  X-rays and radio.
Momentum transfer becomes ineffective at larger distances to drive a wind: a second excitation 
peak in coronal gas is seen at $\sim$150~pc south of the nucleus, coinciding with the negative 
velocity channels. The gas velocities are, however,  of the same magnitude as those of the galaxy disk 
at that location. We thus conclude that the radius of influence of the jet to drive a wind is
restricted to distances not larger than 50~pc $-$ the size of the observed central coronal gas emission.  
On these basis, the estimated bulk kinetic power transferred by the jet to the ISM within the central 
50~pc is $\sim 2 \times 10^{41}$ ($f$/0.1) erg\,s$^{-1}$,  with $f$ being the filling factor, or $\sim$ 6\% 
of NGC\,1386 radiative bolometric luminosity. If the filling factor is close to 1, which is 
probably the case taking into account the strength of the coronal emission and its spatially-resolved 
morphology, the transferred  mechanical power may then surpass the radiative luminosity of the AGN. 
This in turn implies that low Eddington sources as NGC\,1386, although  poor radiators, can however 
be efficient droppers of kinetic energy in to the ISM.
 
When compared to more powerful Seyfert-type AGN for which a similar coronal line energy budget has been 
produced \citep{mue11+}, their ratio of kinetic to bolometric energy is  slightly lower by 
few percent up to an order of magnitude. Their Seyfert sample has bolometric luminosities one to two order 
of magnitude higher than that of NGC\,1386, but the inferred kinetic power is also higher by factor 3 
to an order of magnitude. The way the kinetic power is derived by these authors differs from ours in
two assumptions: {\it (i)} a filling factor 2-dex lower than the one used here;
{\it ii} the bulk of outflow mass moving at the maximum observed 
velocity is about one order of magnitude higher than it is observed as predicted by their biconical 
outflow model. In this way, the mass outloaded into the medium may be overestimated. Therefore, the 
inferred kinetic power in this sample of powerful AGN should be taken as an upper limit. 
In the present case, the remarkable well defined shell morphology of the outflowing gas traced by 
its constant velocity with radius, makes the present estimate robust and independent of model
assumptions.

The resulting mechanical power in the shells represents, in turn, a lower limit to the total 
mechanical power of the putative jet. Assuming a 10\%  efficiency in the transfer of jet power 
to the medium, the jet kinetic power may rise to 10$^{43}$~erg\,s$^{-1}$.

Our scenario is consistent with the kinematics of the optical ionized gas presented in 
\citet{lena15+}. The 2D kinematics of [N\,{\sc ii}]\,$\lambda$6584\,\AA\ discussed by 
these authors is decoupled into 3 components that we interpret as follows: a rotating disc 
in the plane of the galaxy, which we associate with the H$_2$ disc discussed 
in Sect.~\ref{sec:mole_kine}; a broad component, associated with strong nuclear line 
emission. It is confined within 1$\arcsec$ from the 
nucleus, but shows some extension out to approximately 2$\arcsec$. This component is the
one we associate to the expanded shells of 
ionized gas; a third component that involves rotation and/or outflow and extends to 
approximately 2$\arcsec - 3\arcsec$ ($\sim$\,200 pc) either side of the 
nucleus. We associate it to the extended southern tongue of high-ionized gas.

The outflowing gas is expected to be accelerated by the jet-driven shocks. 
This scenario is supported by the strong the coronal emission. Photoionization 
models, where the main source of ionization is the central source, are unable to
explain the observed ratios [Si\,{\sc vi}]/Br$\gamma$ and [Ca\,{\sc viii}]/Br$\gamma$. 
Indeed, no emission at all is predicted at distances larger than a few tens of parsecs, 
in contrast to the observations. The jet-driven shocks could also be responsible
for the observed extended X-rays emission. Therefore, independent 
evidence of a shock-driven NLR is inferred from the analysis of the gas excitation.

\section{Summary} \label{sec:final}

This paper reports the first detection of powerful mass outflows in a low luminosity AGN, 
$L_{\rmn bol} \sim~ 10^{42}$\,erg\,s$^{-1}$. The outflow is inferred from the coronal line gas, specifically 
the [Si\,{\sc vi}] and the [Ca\,{\sc viii}] lines,  which are spatially and kinematically 
resolved within the central 150~pc down to $\sim$15~pc from the centre.  The outflow has the 
shape of a two expanding ionized gas shells moving in opposite directions relative to the 
nucleus and along our line of sight.

We propose  that these expanding shells are 
driven mechanically by the NGC\,1386 incipient core jet. The kinematics of the ionized gas 
is distinctive from that of the molecular H$_2$, and of Br$\gamma$ to a lesser extend, 
which shares regular rotation with the stellar component albeit in a different disc inclination.
The lack of rotation in the coronal gas, in addition to the very small gradient
in velocity measured between the edges of the shells indicate that we are seeing a swept 
layer of material accelerated by the  radio-jet. The shocks produced by the jet into the 
medium account for the ionization/excitation of the coronal gas from the 
nucleus up to $\sim$120\,pc. Nuclear photoionization is found too weak to explain the
size of the emission region and strength of the high-ionization gas. Shocks furthermore 
also account for the nuclear H$_2$ line 
ratios measured. We propose that the same mechanism accounting for the extended soft X-ray 
emission that encloses the expanding shells is also responsible for the radio emission 
that is co-spatial with the receding expanding shell. 

The morphology and location of the shells at a few tens of parsec from the centre allows us 
to get a first-order estimate of the outflow mass rate of $\sim$11~M$_\odot$~yr$^{-1}$, which is 
among the largest estimated in counterpart powerful AGNs. We get a tentative inflow rate 
from the H$_2$ gas of $< 0.4$~M$_\odot$~yr$^{-1}$, more than an order of magnitude lower 
than the outflow rate. This result is common to other AGN, implying that the outflowing material 
comes from the nuclear surroundings.  We find that the momentum transfer 
of the jet to drive the outflow is limited to a radius of influence of $<150$~pc from the 
centre as we show that the coronal gas is already circularized to the velocities 
of the galaxy disk at those locations.

The kinetic energy deposited by the outflow into the medium is $\geq 10^{41}$~erg~s$^{-1}$, 
$\geq 6$\% of the bolometric luminosity. An  order of magnitude higher is predicted if 
the gas filling factor is close to 1, which it may well be the case taking into account 
the strength and uniformity of the coronal emission in the shells.  Models  of quasar 
evolution require about 5\% of L$_{\rmn bol}$  to explain the BH mass -- bulge mass 
correlation. We show that this energy boundary is achievable by a galaxy  3 to  4 orders 
of magnitude less powerful -- radiatively --  than a quasar. 
Due to its condition of early type galaxy, in galaxy evolution scenarios, NGC\,1386 has presumably 
passed its quasar phase and is now getting in the later activity stages, which is usually 
characterized by a Low-luminosity AGN. 
This work shows that low Eddington sources as NGC\,1386, although poor radiators, can 
however be efficient droppers of kinetic energy into the ISM.

\section*{Acknowledgments}

We are grateful to the anonymous Referee for helpful suggestions to 
improve this manuscript.
{\it A.R-A.} acknowledges the Conselho Nacional de Desenvolvimento Cient\'ifico
e Tecnol\'ogico, CNPq. through grant  311935/2015-0) and to the
Severo Ochoa Program for partial support to this work. {\it F. M-S.}
acknowledges financial support from NASA HST Grant HST-AR-13260.001.

\label{lastpage}

\end{document}